\documentclass[10pt]{iopart}
\usepackage[normalem]{ulem}
\usepackage{multirow}
\usepackage[utf8]{inputenc}
\usepackage[authoryear]{natbib}
\usepackage{iopams}
\usepackage{hyperref}
\usepackage{color}  
\expandafter\let\csname equation*\endcsname\relax
\expandafter\let\csname endequation*\endcsname\relax
\usepackage{mathtools}
\usepackage[]{graphicx}
\usepackage{multicol}
\usepackage{cooltooltips}
\usepackage{pdfcomment}
\newcommand{\iotab}{\lower3pt\hbox{$\mathchar'26$}\mkern-7mu\iota}

\usepackage[colorinlistoftodos]{todonotes}
\graphicspath{{./}
	{./figs/png/}
}

\begin{document}

\title[]{CIEMAT-QI4X: a reactor-relevant quasi-isodynamic stellarator configuration  compatible with an island divertor}
%
\author{E. Sánchez, J. L. Velasco, I. Calvo, J. M. García-Regaña, C. Salcuni and J. A. Alonso} 

\address{
Laboratorio Nacional de Fusión, CIEMAT, 28040, Madrid, Spain
}

\ead{edi.sanchez@ciemat.es}
\vspace{10pt}

\begin{abstract}
	
	A four-field-period quasi-isodynamic stellarator configuration is presented that exhibits small neoclassical and electrostatic turbulent transport, good fast-ion confinement over a wide range of $\beta$ values, small bootstrap current and an edge island structure compatible with an island divertor. This configuration, called CIEMAT-QI4X, has been obtained by building on the optimization strategy and sophisticating the methods employed in [Sánchez E. et al 2023 Nucl. Fusion 63 066037]. The optimization has been improved by incorporating metrics to control Mercier stability and by enforcing strict constraints on the rotational transform profile to achieve nested toroidal surfaces in the confinement region and a divertor island structure at the plasma edge. Specifically, CIEMAT-QI4X has a 4/4 island chain at the edge that is resilient at least up to $\beta=4\%$, even when the bootstrap current is included. A corresponding set of filamentary coils is presented that generates the configuration with enough accuracy to preserve the aforementioned physics properties. In terms of physics performance, CIEMAT-QI4X establishes as a candidate for a stellarator fusion reactor design.

\end{abstract}
%

\ioptwocol

%
\section{Introduction}
The stellarator concept \cite{Helander2012a} offers important advantages over the tokamak for a fusion reactor. The absence of inductive current makes stellarators less prone to current-driven instabilities and disruptions, and facilitates steady-state operation. However, unlike in tokamaks, good confinement of collisionless particles and energy is not guaranteed in a generic stellarator  \cite{galeev1979review,beidler_benchmarking_2011, Calvo2013a, calvo_effect_2017} and requires careful magnetic field design, usually called stellarator optimization.

In order to achieve good confinement of collisionless particle orbits, stellarator optimization strategies have typically been based on designing approximately omnigenous fields. Omnigenity \cite{cary_omnigenity_1997} guarantees that the net radial drift of all particle orbits vanishes, like in a tokamak.
Two main classes of omnigenous fields can be distinguished: quasi-symmetric (QS) \cite{Boozer1995,Nuhrenberg1988,garabedian_stellarators_1996} and quasi-isodynamic (QI) fields \cite{nuhrenberg_development_2010}, In QS configurations, the magnetic field strength exhibits a symmetry along either the toroidal, poloidal, or helical direction (in Boozer coordinates \cite{BoozerCoord}). In QI configurations, by contrast, the magnetic field strength contours close poloidally without an explicit symmetry. A key advantage of the QI concept is that a perfectly QI field has identically zero bootstrap current \cite{subbotin_integrated_2006, Helander2009, LandremanM2012}, ensuring that the rotational transform profile is not modified by plasma self-driven currents. Fields that deviate only slightly from perfect quasy-isodynamicity still exhibit very small bootstrap currents, providing resilience of the rotational transform profile with plasma pressure. This, in turn, has a strong positive impact on the existence of a stable magnetic island at the plasma edge, which can be exploited to design a resonant divertor, which helps exhausting particle and heat from the reactor \cite{grigull_first_2001,renner_divertor_2002}. 
New strategies where the target magnetic field is far from omnigenous are also being explored  \cite{velasco_piecewise_2024, velascoExplorationParameterSpace2025} and a family of these new optimized configurations, which exhibit low bootstrap current and could then be compatible with an island divertor, has been found \cite{calvo_piecewise_2025}. 

Significant progress in stellarator optimization has been made in recent years, and new configurations have been obtained approaching either QS  \cite{landreman_magnetic_2022} 
 or QI  \cite{goodman_constructing_2023} fields with high precision. 
 Recent optimizations have targeted not only improved confinement of bulk ions, already optimized and experimentally demonstrated successful  in W7-X \cite{Grieger1992}, but also improved fast ion confinement and reduced turbulent transport, which are aspects for which the W7-X configurations are not sufficiently optimized. Many new configurations have been obtained  \cite{bader_advancing_2020,henneberg2019,sanchez_quasi-isodynamic_2023,goodman_constructing_2023,bindelDirectOptimizationFastIon2023, garcia-regana_reduced_2024,goodman_quasi-isodynamic_2024,kim_optimization_2024,lionStellarisHighfieldQuasiisodynamic2025, hegnaInfinityTwoFusion2025}
	 that improve upon the W7-X stellarator  in some or all these respects. 

Recent optimization studies of QI (and QS)  configurations often assume the presence of well-formed flux surfaces but give insufficient consideration to this aspect during the optimization process.
 Although well-nested flux surfaces are essential for any toroidal confinement configuration, they are typically not treated as an explicit objective in optimization. 
The quality of flux surfaces is closely related to the rotational transform, because at the radial positions where the rotational transform takes a rational value of the form  $\iota=\frac{n}{m}$, with $n$  and $m$ integers \footnote{When $n$ is a multiple of the device periodicity, the island is called natural, which are the most deleterious ones.} flux surfaces can be destroyed and islands can form due to resonant magnetic field perturbations.  
The island width, $w$, depends on the order, $m$, and magnetic shear, $\iota\prime=\frac{d\iota}{dr}$, as  \cite{boozer_non-axisymmetric_2015}
\begin{equation}
	w = 2 \sqrt{\frac{R B_{mn}}{m \iota^\prime}},
	\label{Eq:islandWidth}
\end{equation} 
where $R$ is the major radius and $B_{mn}$ denotes the amplitude of the radial Fourier components of the magnetic field strength (normalized to a reference value, such as the magnetic field at the magnetic axis) that resonate with the island numbers $m,n$. 
Then,  the lower the order of the island, $m$, the larger its size and, consequently, the more detrimental its effect on confinement. 
The resonant components $B_{mn}$ can arise either from the coil design itself or from manufacturing and positioning errors.
Procedures for designing modular coils that minimize specific resonant components  $B_{mn}$ have been proposed \cite{hudson_eliminating_2002, zhu_identification_2019, thomas_kruger_minimizing_2022} and in  ref. \cite{smiet_efficient_2025}, the reduction of the island size is performed in a single-stage optimization for a rotating ellipse equilibrium. 
 It can be shown that in low magnetic shear configurations, even very small perturbations ($B_{mn}\approx 10^{-4}$) can generate islands of significant width around low-order rational values of $\iota$, which underscores the need for extremely precise design, construction, and alignment of the coils, especially when the magnetic shear is small in the vicinity of a low-order rational value (see, e.g. \cite{pedersen_confirmation_2016, neilson_lessons_2010, zhu_identification_2019, lobsien_stellarator_2018}).  

The lowest-order islands can be avoided by selecting $\iota$ values sufficiently far from the lowest-order rationals, which is standard practice in the optimization of QI configurations. Typically, $\iota$ is constrained during optimization to prevent such rationals from lying within the last closed flux surface. However, some higher-order rationals are unavoidable. If a strict control of the entire profile is not imposed, as it is usually the case \cite{sanchez_quasi-isodynamic_2023, goodman_constructing_2023, goodman_quasi-isodynamic_2024, dudt_magnetic_2024}, islands associated to higher-order rationals can still form in the plasma core with significant size at locations with very low magnetic shear. 
In some cases, 
the rotational transform profile is even less constrained, it is allowed to evolve freely, or only weakly constrained, during optimization \cite{bader_advancing_2020, landreman_magnetic_2022} and the bootstrap current can significantly reshape it further. Anticipating the bootstrap current contribution during the design phase to ensure  good nested flux surfaces at finite-$\beta$ \cite{ku_new_2006} is difficult, because the bootstrap current depends strongly on the density and temperature profiles.

The overlap of neighboring islands can degrade the quality of flux surfaces, creating ergodic regions \cite{ohyabuLargeHelicalDevice1994}. This effect becomes more pronounced with large magnetic shear: the separation between adjacent islands scales inversely with the shear, while their width scales with the inverse square root of the shear, as given by Eq.~(\ref{Eq:islandWidth}). Consequently, island overlap can be mitigated by maintaining sufficiently low shear. An optimal shear value must therefore be identified to balance island size and overlap.

Finally, when an island divertor is sought, $\iota$ at the plasma edge is usually constrained to approach a low-order rational, such that an edge island forms outside the last closed flux surface and can serve as the divertor \cite{grigull_first_2001, feng_physics_2006, okamura_island_2020,renner_divertor_2002}. But this is not enough, because the location and size of this island depend not only on the edge $\iota$ but also on the edge magnetic shear. If the shear is too large, the divertor island becomes too small; if too small, and $\iota$ is  close to a rational, the edge island may become excessively wide, reducing the effective plasma volume. 

In this work, we demonstrate how the impact of magnetic islands in a QI stellarator can be mitigated by enforcing strict control of the rotational transform profile during the optimization. This approach improves confinement, stabilizes the edge island, and enhances resilience to coil construction and alignment errors. 
We present a new configuration from the CIEMAT-QI family \cite{sanchez_quasi-isodynamic_2023, velasco_robust_2023}, called CIEMAT-QI4X, which has been obtained using this strategy and combines desirable physics properties: MHD stability, good neoclassical transport, reduced bootstrap current, good fast-ion confinement, and reduced turbulence. Moreover, it features a rotational transform profile that ensures robust flux surfaces in the plasma core while supporting a sufficiently large and stable divertor island. 
A set of modular, in-principle buildable, filamentary coils generating this configuration is also provided.

The paper is organized as follows. Section~\ref{secOptimFSID} describes the strategy for optimizing flux surfaces and the island divertor. Section~\ref{secConfigProps} presents the physics properties of the optimized configuration. 
Finally, Section~\ref{secSumandConc} summarizes the main results and provides conclusions.

\section{Optimization of flux surfaces and edge island structure}\label{secOptimFSID}
%

In this section, we show how, starting from the CIEMAT-QI4 configuration \cite{sanchez_quasi-isodynamic_2023}, we obtain a new configuration in which the width and overlap of low-order islands within the plasma column are reduced, while the structure and extent of the edge island chain are improved. These improvements are achieved without significantly degrading the physical properties of the original  configuration.

\begin{figure}
	\centering
	\includegraphics[draft=false,  trim=0 10 0 0, clip, width=8.9cm]{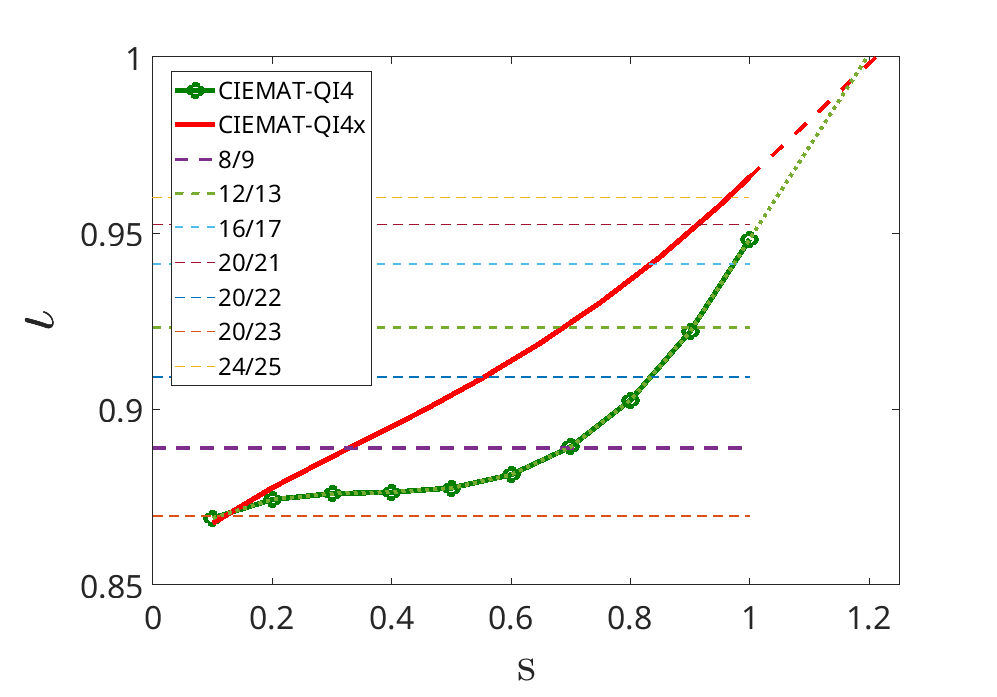}
	\caption{Rotational transform profiles for the CIEMAT-QI4 (green) and the CIEMAT-QI4X (red) optimized configurations. Dashed lines are shown for the values of the lowest-order rational values crossed by the profiles. The $\iota$ profiles are extrapolated outside the last closed flux surface until they reach the $\frac{4}{4}$ rational, which is pursused to form an island divertor structure.}
	\label{fig:rotationalTransforms}
\end{figure}
 \begin{figure*}
 	\centering
 	\includegraphics[draft=false,  trim=390 285 330 220, clip, width=7.95cm]{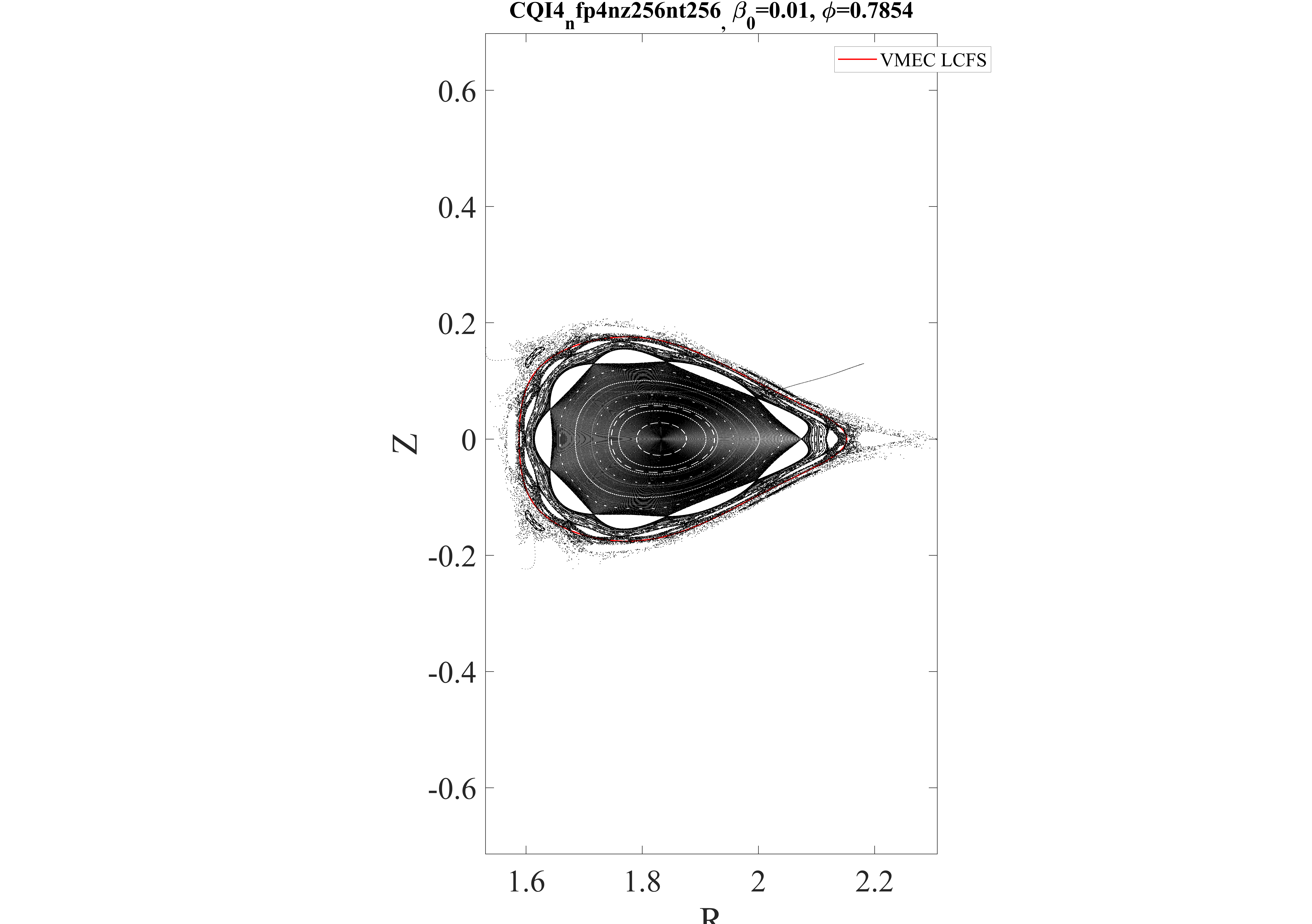}
 	\includegraphics[draft=false,  trim=300 310 200 300, clip, width=7.95cm]{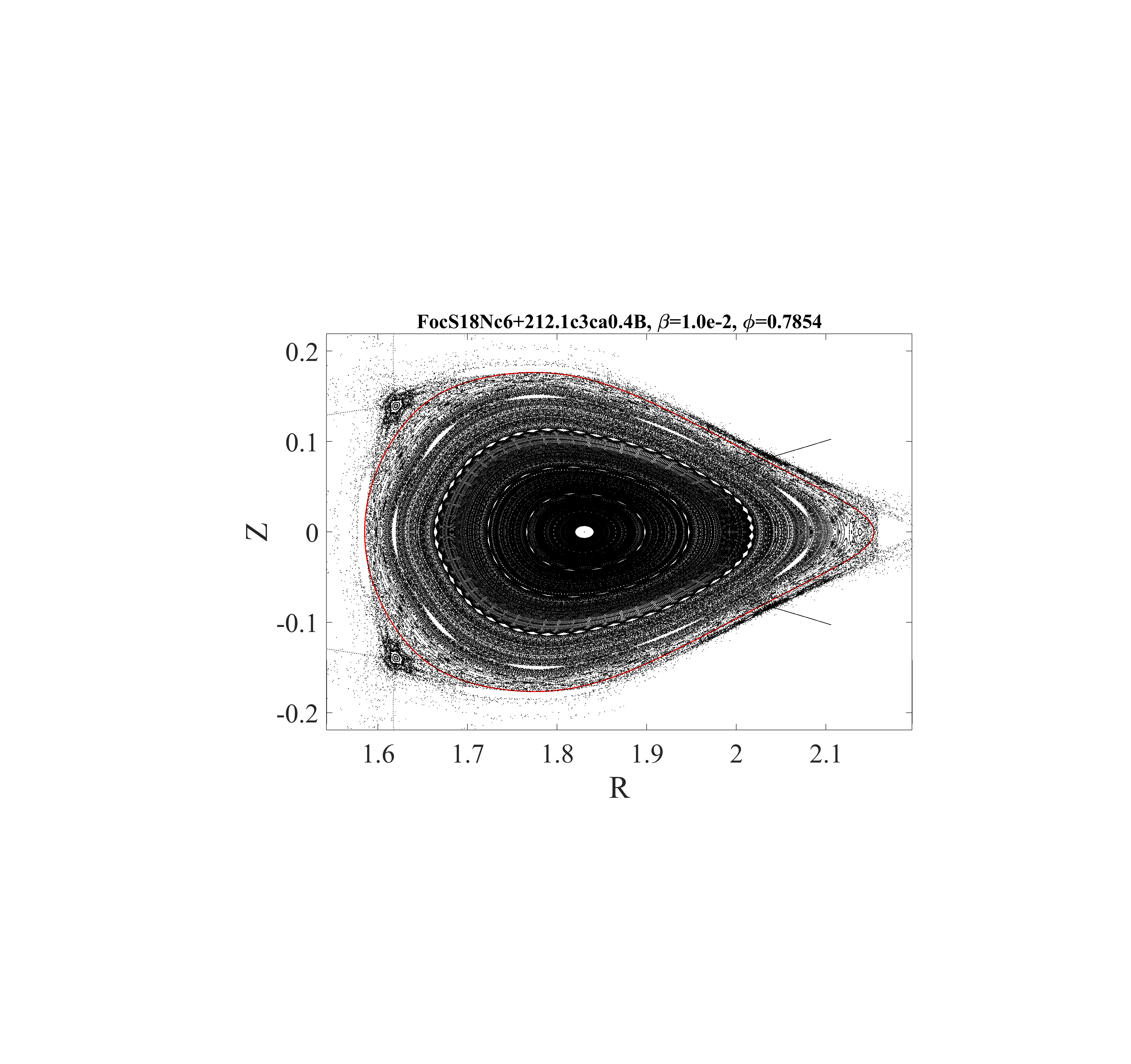}
 	\caption{Poincaré plots at toroidal angle $\phi=\pi/4$ for the CIEMAT-QI4 configuration at $\beta=0.5\%$ obtained with HINT-3D calculations, using an optimized set of coils producing the configuration with good fidelity (left) \cite{SanchezEFTC23} and with a coil set optimized to minimize the $\frac{8}{9}$ island width (right) \cite{sanchezSimons24}. The last closed flux surface for a VMEC calculation at same $\beta$ value is shown in red for reference. The $\frac{8}{9}$~, $\frac{12}{13}$, and $\frac{16}{17}$  islands are seen in the plasma column with some island overlapping at the plasma edge in the left case.}
 	\label{fig:cqi4Poincare}
 \end{figure*}

Figure~\ref{fig:rotationalTransforms} shows the rotational transform profiles for the original CIEMAT-QI4  and the new CIEMAT-QI4X configuration, whose other details will be presented in Section~\ref{secConfigProps}. The profiles are extrapolated until they reach $\iota=1$ (for $s>1$), corresponding to the location of the $\tfrac{4}{4}$ island pursued for forming a divertor structure\footnote{Note that the radial coordinate $s$ is not well defined outside $s=1$.}. Here, $s$ denotes the normalized toroidal flux, used as radial coordinate. The lowest-order rational in the CIEMAT-QI4 profile is $\tfrac{8}{9}$, which is crossed around $s \approx 0.7$.  

In figure~\ref{fig:cqi4Poincare}, two Poincaré maps at toroidal angle $\phi=\pi/4$ are shown for the CIEMAT-QI4 configuration at $\beta=0.5\%$, computed with HINT-3D using two coil sets that reproduce the configuration with high fidelity \cite{SanchezEFTC23,sanchezSimons24}. The last closed flux surface for two free boundary VMEC equilibria using the same coils and for the same $\beta$ value are included for reference. In  figure~\ref{fig:cqi4Poincare}-left, the $\tfrac{8}{9}$ island is observed in the plasma core, accompanied by higher-order islands such as $\tfrac{12}{13}$ and $\tfrac{16}{17}$ at the plasma edge, with some overlap between them. The island widths depend on the coil design and vary slightly with $\beta$, but the $\tfrac{8}{9}$ island consistently appears  with significant size for many coil designs.  
The natural $\tfrac{4}{4}$ island, intended to provide a divertor structure, is also present outside the last closed flux surface. However, its size and topology—both of which vary with $\beta$—are not suitable for a divertor.

Optimized coil designs for the CIEMAT-QI4 configuration were obtained with dedicated tools in the FOCUS code \cite{zhu_identification_2019,thomas_kruger_minimizing_2022}, which minimize radial magnetic field components resonating with the $\tfrac{8}{9}$ island, thus effectively reducing the $\tfrac{8}{9}$ island width, as shown in figure \ref{fig:cqi4Poincare} right for one of these coil sets. 
However, simultaneously suppressing multiple islands and reducing the overlap is substantially more difficult. Moreover, these methods target only the closed-flux-surface region and cannot improve the divertor island. Relying exclusively on this approach is risky: while millimetric coil adjustments may reduce island size significantly, equally small construction or positioning errors can instead amplify the islands—even if the coils are carefully optimized to suppress them.

We adopt a different strategy by imposing a much stricter control of the rotational transform profile during the optimization process than is usually done. As discussed in the introduction, simultaneously reducing the size of low-order islands and their overlap requires balancing two conflicting objectives: increasing the magnetic shear reduces island width, whereas decreasing it helps to mitigate island overlap. Achieving an appropriate compromise between these goals is therefore essential. Furthermore, the formation of a significant edge island, required for a divertor, depends also critically on the edge rotational transform profile. Although the control of the rotational transform profile constrains the magnetic field, can influence the physical properties of the configuration and may even conflict with other optimization criteria, its benefits are fundamental and of paramount importance.	
	Controlling the rotational transform profile during optimization—typically carried out in vacuum or at low $\beta$-- defines a baseline profile that will inevitably be modified at higher plasma pressure through nonlinear processes, particularly by self-generated currents such as bootstrap. In a near-quasi-isodynamic configuration, however, this should not pose a major problem, since the bootstrap current is expected to be very small.
	
	A strict control of the rotational transform profile was imposed along the optimization process with the goal of increasing the magnetic shear around the location of $\iota=\frac{8}{9}$  and reducing it at the plasma edge region to avoid island overlap (see figure \ref{fig:rotationalTransforms}). This is done by keeping the innermost and outermost target rotational transform values close to those of the CIEMAT-QI4 profile and setting additional target values for $\iota$, with tight tolerances at a set of intermediate  radial locations. Beyond these constraints on $\iota$, a new target of magnetic shear has been introduced. 
	The complete list of targets used in this optimization process includes, in addition to the rotational transform profile, the magnetic shear, and the Mercier criterion, $D_M$ (see section \ref{secConfigProps}), as well as all the targets previously employed in the CIEMAT-QI4 configuration: effective ripple ($\epsilon_{eff}$), magnetic well (W), and $\Gamma_c$, $\Gamma_{\alpha}$, VBT, VBB, and WBW, which are used to approach QI and maximum-$J$ properties (see Ref. \cite{sanchez_quasi-isodynamic_2023} for details). 
	\begin{figure*}
		\centering
		\includegraphics[draft=false,  trim=360 0 380 5, clip, width=7.5cm]{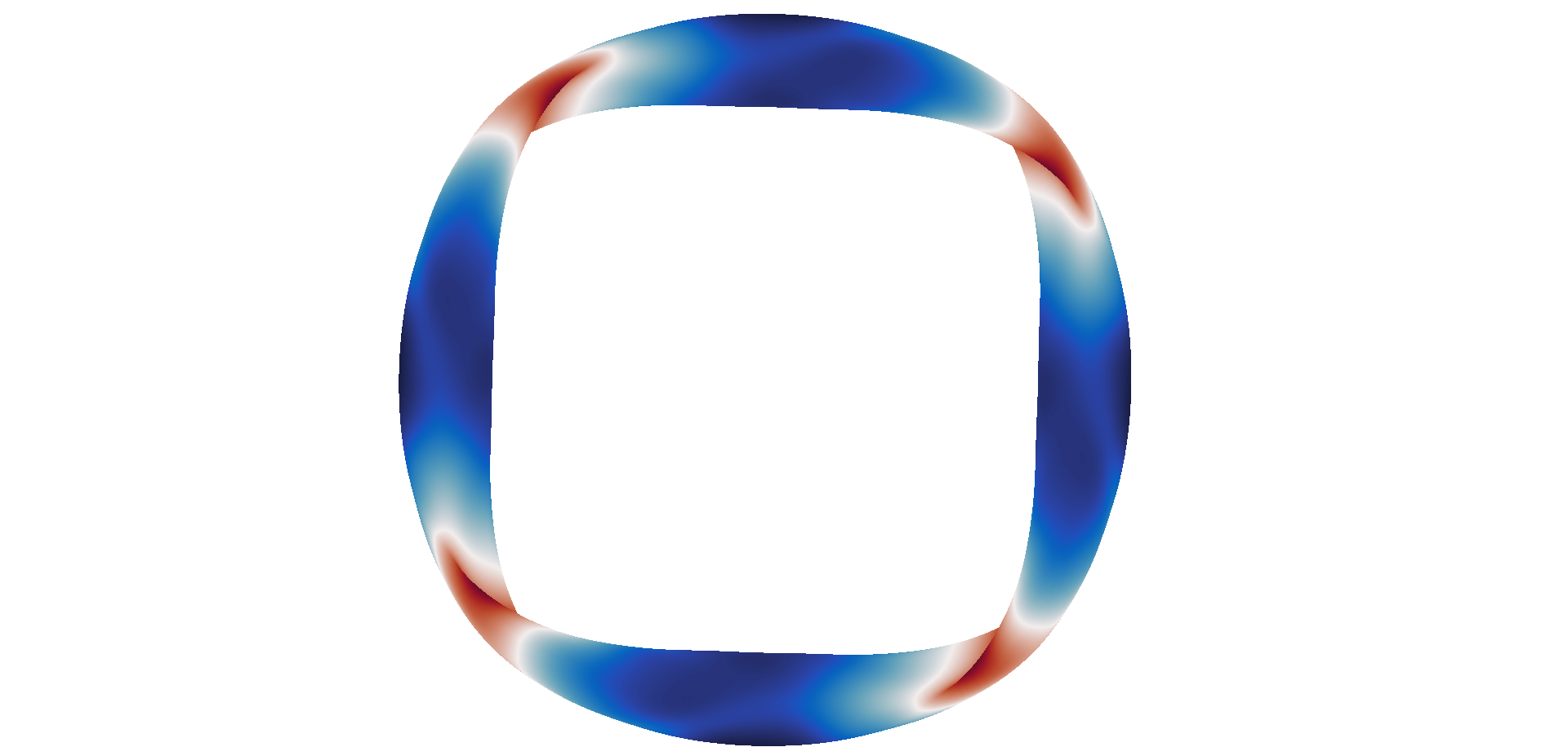}
		\raisebox{-0.0\height}{
			\includegraphics[draft=false,  trim=0 40 0 80, clip, width=7.5cm]{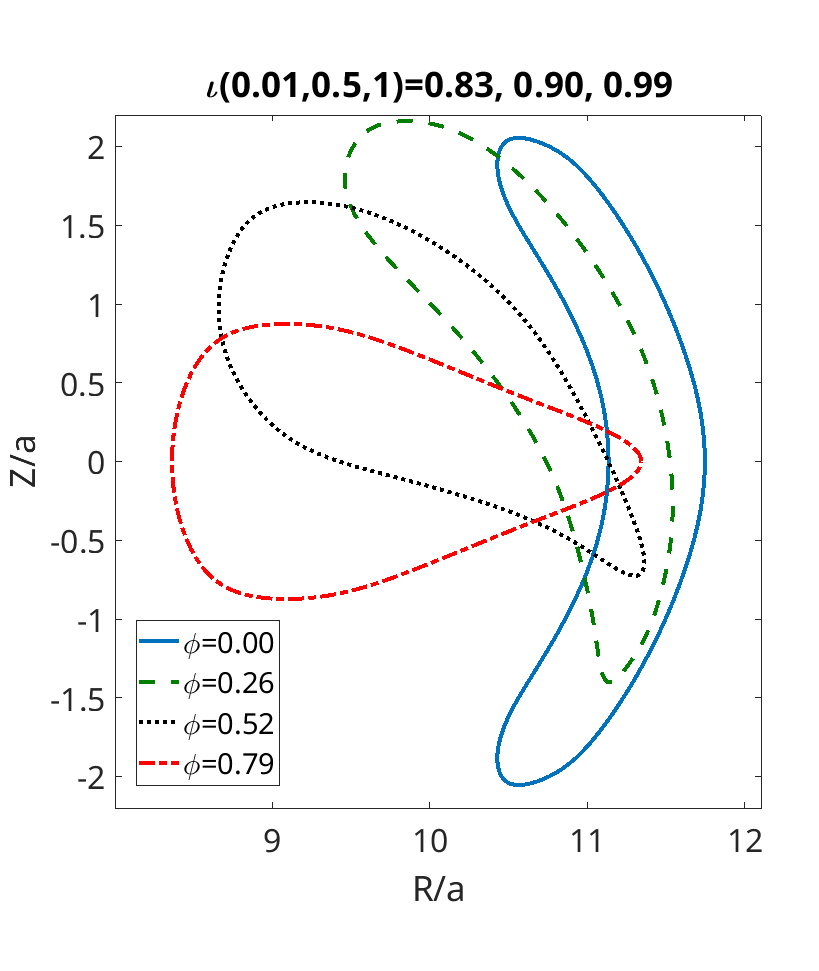}
		}
		\caption{Magnetic field strength at the last closed flux surface (left) and Poincaré plots for the CIEMAT-QI4X configuration at several toroidal angles (right)}
		\label{fig:BStrengthyPoincare}.
	\end{figure*}	
	It should be noted that a QI field is tightly linked to its rotational transform profile, and consequently, modifying the latter can strongly affect its degree of omnigenity. For this reason, changes to the rotational transform profile in a QI configuration require re-optimization to preserve omnigenity. An iterative trial-and-error process was therefore followed until an appropriate $\iota$ profile, providing a good balance among all optimization targets, was obtained.
	
	As a result, a new configuration, called CIEMAT-QI4X, was identified. This configuration features a rotational transform profile (shown in figure \ref{fig:rotationalTransforms}) that minimizes the size of the islands inside the plasma and their overlap, and improves the divertor island chain. 
	In the next section the CIEMAT-QI4X configuration and its physics properties are described in detail.

	 %
	 \section{The CIEMAT-QI4X configuration}\label{secConfigProps}
	
	As discussed in the previous section, the CIEMAT-QI4X configuration was derived from CIEMAT-QI4 \cite{sanchez_quasi-isodynamic_2023} and retains similar global properties. It is a four-period quasi-isodynamic configuration with aspect ratio $A=10.4$. At $\phi=0$, the plasma elongation is $\kappa = \tfrac{\Delta Z}{\Delta R} = 6.9$, where $\Delta R$ and $\Delta Z$ denote the radial extent in the $R$ and $Z$ directions, respectively. The magnetic field strength on the last closed flux surface of CIEMAT-QI4X, together with Poincaré plots at several toroidal angles, are shown in figure~\ref{fig:BStrengthyPoincare}.  
	
	Since many physics properties presented in this section are evaluated for free-boundary equilibria taking into account the coils, we begin this section by presenting the filamentary coil set designed for this configuration.
	
	\subsection{Filamentary coils for CIEMAT-QI4X}\label{secCoils}
	The first design of coils for CIEMAT-QI4, published in Ref. \cite{sanchez_quasi-isodynamic_2023}, was obtained with a minimum coil–plasma distance comparable to the minor radius of the configuration. Several optimized coil designs were later generated for that configuration using the FOCUS code \cite{zhu_new_2018}, varying the minimum coil–plasma distance. These optimizations produced coil sets with improved coil–coil distance, reduced coil complexity, and lower maximum curvature and torsion \cite{sanchezSimons24,SanchezEPS23,SanchezEFTC23}.  
		
	\begin{figure*}
		\centering
		\includegraphics[draft=false,  trim=190 20 100 20, clip, width=19.15cm]{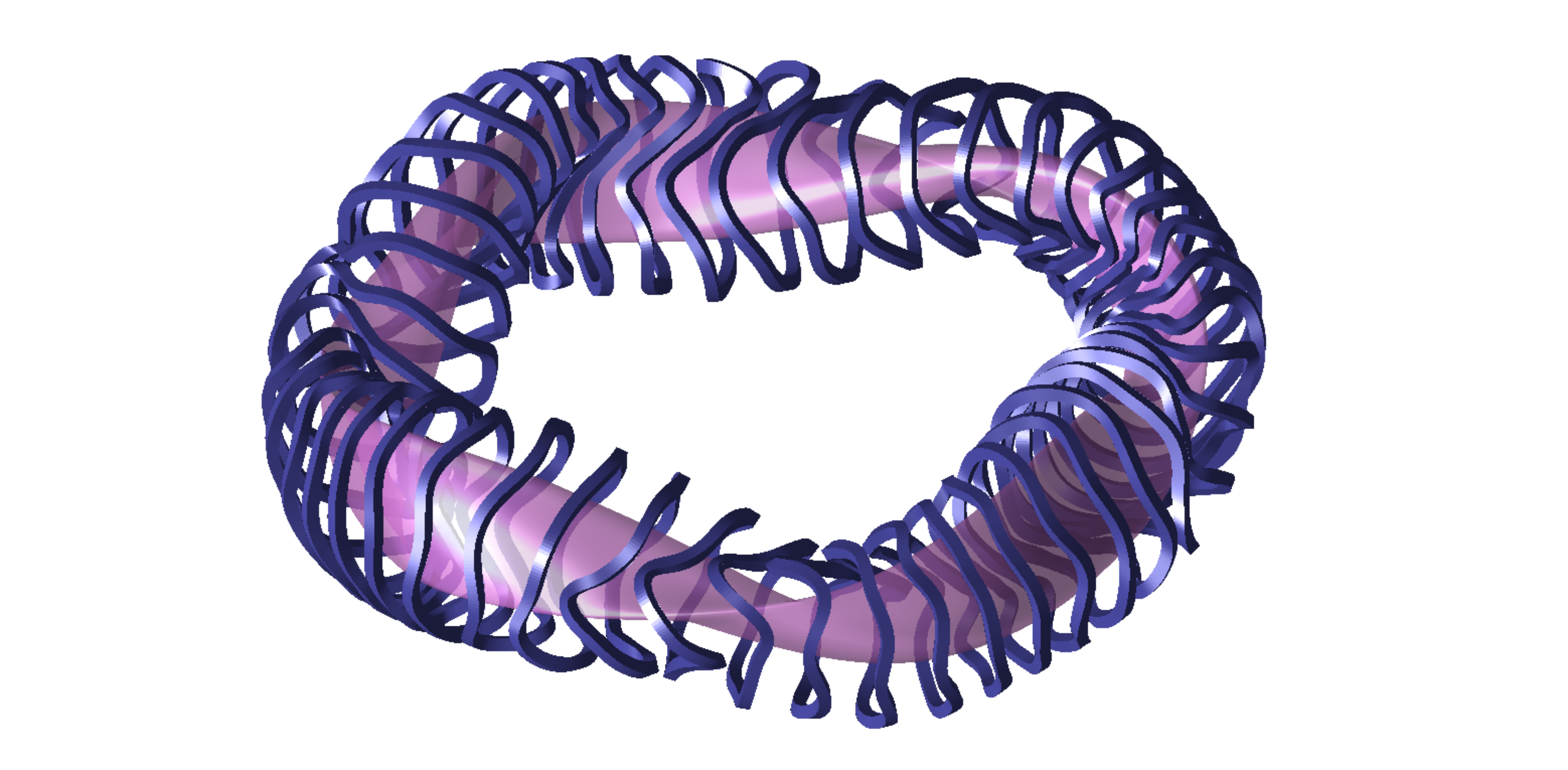}
		\caption{Modular coils designed for the CIEMAT-QI4X configuration. The coils are filamentary and the volume is only shown for visualization purposes. This coil design is the one producing the flux surfaces in figures \ref{fig:CQI4XPoincare} and the results shown in section \ref{FIMetrics}  }
		\label{fig:filamentaryCoils}
	\end{figure*}
	First studies of a breeding blanket for CIEMAT-QI4 assessed the minimum coil–plasma distance required to accommodate a breeding blanket in a reactor-scale device \cite{SanchezEFTC23}. These preliminary results indicated that, for a CIEMAT-QI4 configuration scaled to reactor size ($R \approx 16$ m), the minimum coil–plasma distance could be reduced significantly (below the minor radius) while still allowing a blanket capable of providing a sufficient tritium breeding ratio (TBR).  
	
	Building on these results, we have carried out several coil designs for CIEMAT-QI4X (which closely resembles CIEMAT-QI4) with different minimum coil–plasma distances. This has enabled us to assess how coil complexity, maximum curvature and torsion, and fast-ion confinement depend on the coil–plasma spacing. The general trend, consistent with previous works \cite{kappel_magnetic_2024}, is that coils placed closer to the plasma are simpler, with smaller curvature and torsion. Coil–coil spacing can also be improved by reducing the coil–plasma distance compared with CIEMAT-QI4. However, the coil ripple increases as the coils approach the plasma, which is known to affect neoclassical transport and, in particular, fast-ion confinement.  
	
	Taking all these considerations into account, we have designed a coil set for CIEMAT-QI4X, consisting of 48 coils, with a minimum coil–plasma distance of $d_{\text{min}} \approx 0.84\,a$, where $a$ is the minor radius. At this spacing, we achieve an optimal compromise between coil complexity, coil–coil separation, and magnetic field accuracy. The flux-surface-averaged normal field error is $ 3.5 \times 10^{-3}$, which is sufficiently low to avoid significant degradation of fast-ion confinement (see Section \ref{secConfigProps}). Further reducing this error—at the cost of greater coil complexity—did not improve fast-ion confinement, in agreement with previous findings \cite{wiedman_coil_2024}.
	
	This coil set represents a substantial improvement over those designed for CIEMAT-QI4 in terms of coil complexity, maximum curvature and torsion, and coil–coil distance. The design is shown in figure \ref{fig:filamentaryCoils}, and its main parameters are listed in Table \ref{tablaCoils}.  	
\begin{table}
	\centering
	\caption{Main parameters of the coil set shown in figure \ref{fig:filamentaryCoils}, for a reactor-scaled CIEMAT-QI4X configuration with $R=18.5~m$ and $a=1.77~m$.}
	\label{tablaCoils}
	\vspace{.2cm}
	\begin{tabular}{|l|c|}
		\hline
		number of coils & 48 \\
		average length (m)& 30.7 \\ 
		maximum curvature (m$^{-1}$)& 1.85 \\ 
		average curvature (m$^{-1}$)& 0.38 \\ 
		average torsion (m$^{-1}$)& 0.20 \\
		minimum coil-plasma distance (m) & 1.48 \\
		minimum coil-coil distance (m) & 0.84 \\
		surface-averaged $B_N$ error & $3.5 \times 10^{-3}$\\
		\hline
	\end{tabular}
\end{table}
	This coil design appears, in principle, to be feasible when compared with earlier proposals in the literature \cite{grieger_modular_1992,rise_experiences_2009,schauer_helias_2013}. It is simpler than some previous designs \cite{neilson_lessons_2010} and not qualitatively more complex than more recent ones \cite{lionStellarisHighfieldQuasiisodynamic2025,hegnaInfinityTwoFusion2025}. Detailed feasibility studies with different conductor technologies, including low-temperature superconductors (LTS) and high-temperature superconductors (HTS), are ongoing and will be reported elsewhere.  
	
	Based on the earlier breeding blanket assessment for CIEMAT-QI4, the minimum coil–plasma distance in CIEMAT-QI4X also appears sufficient to accommodate a blanket with adequate TBR. A dedicated evaluation of breeding blanket integration for this configuration is underway and will be presented in a future work.  
	
	We now turn to the magnetic field structure, flux surfaces, and divertor island topology, which constitute the main novelties of CIEMAT-QI4X compared with CIEMAT-QI4.  
	
	\begin{figure*}[!ht]
	\centering
	\includegraphics[draft=false,  trim=315 330 255 365, clip, width=7.5cm]{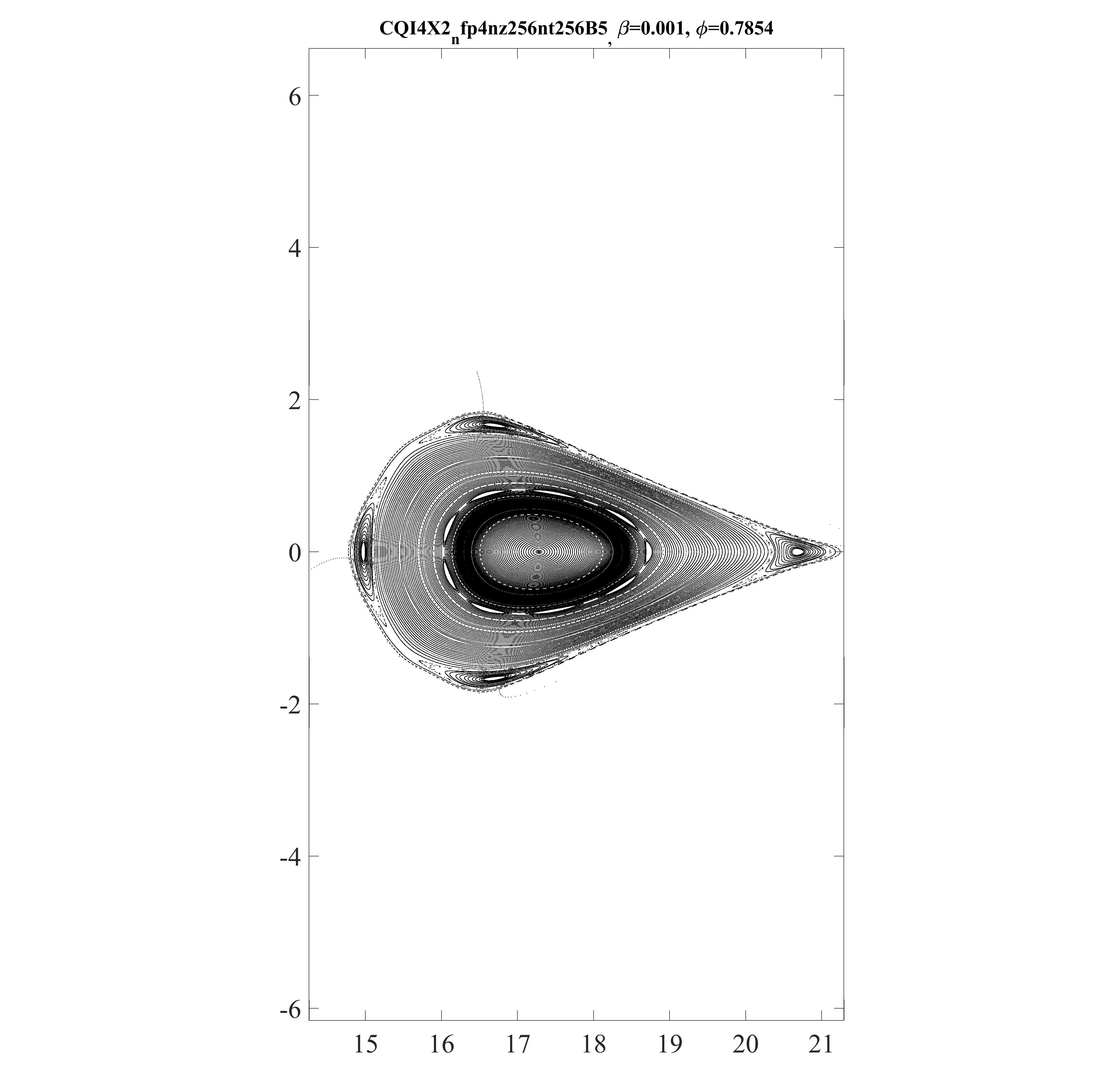}
	\includegraphics[draft=false,  trim=290 100 215 50, clip, width=5.75cm]{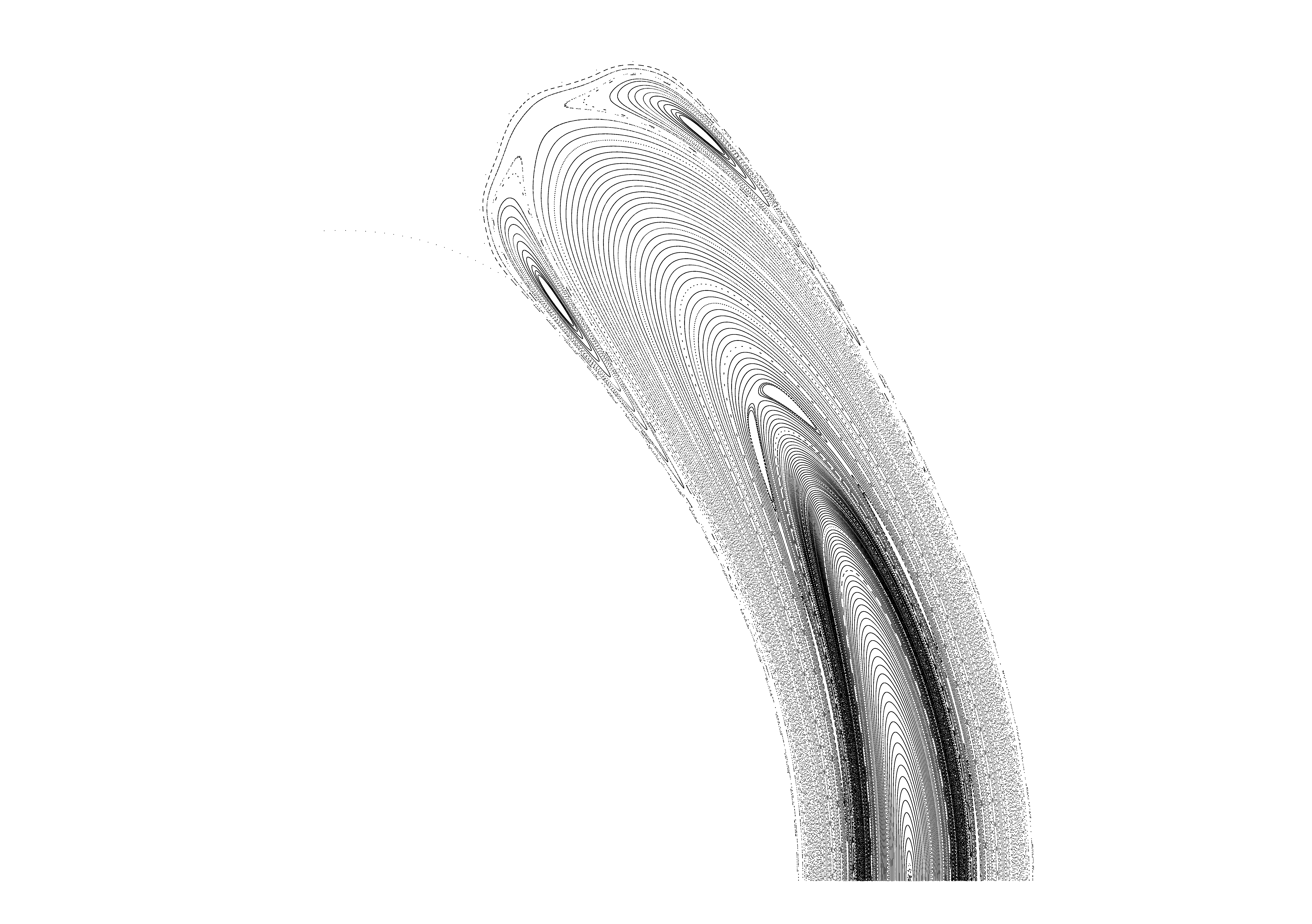}\vspace{0.15cm}
	\includegraphics[draft=false,  trim=315 330 255 365, clip, width=7.5cm]{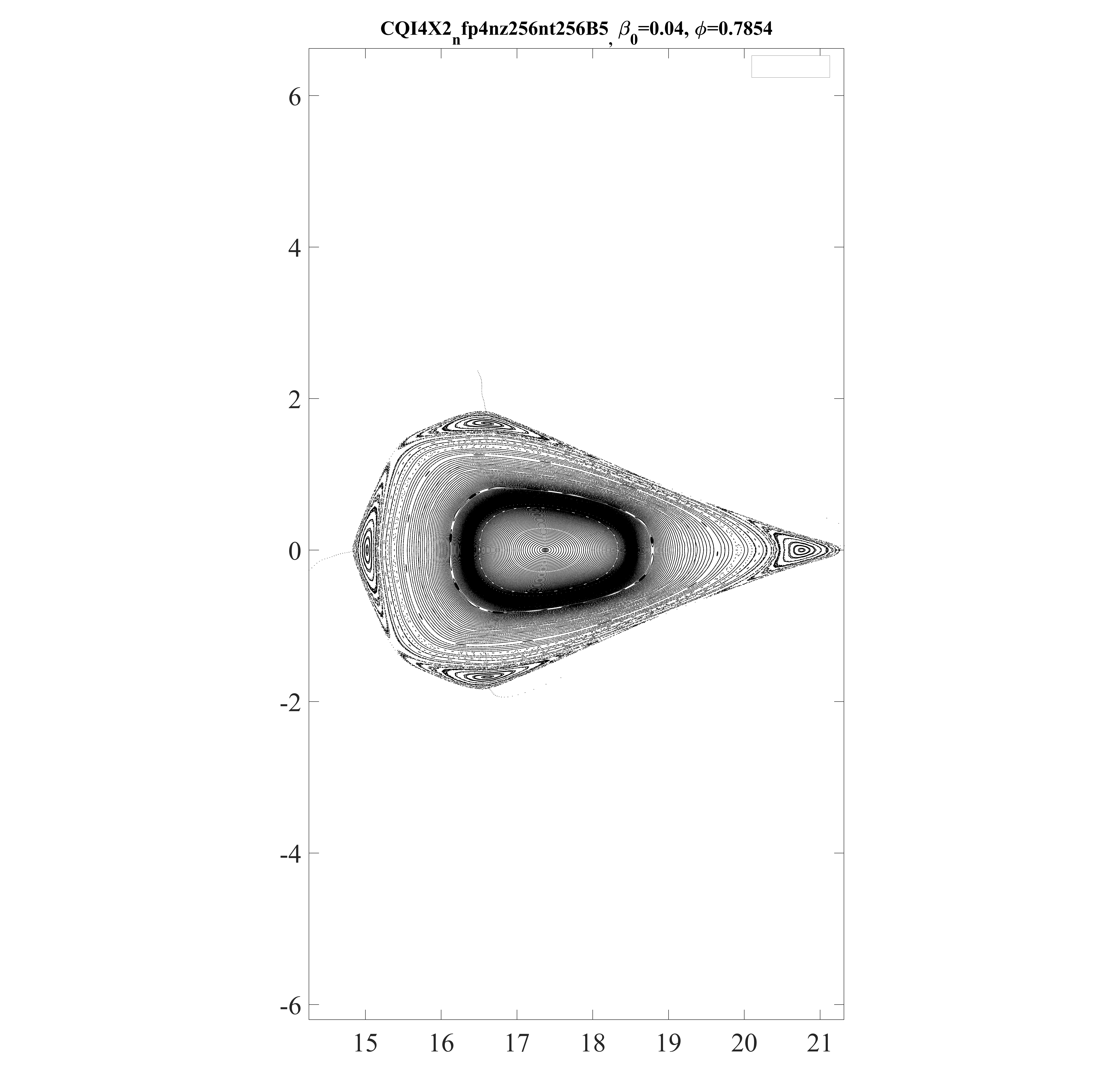}
	\includegraphics[draft=false,  trim=290 100 215 50, clip, width=5.75cm]{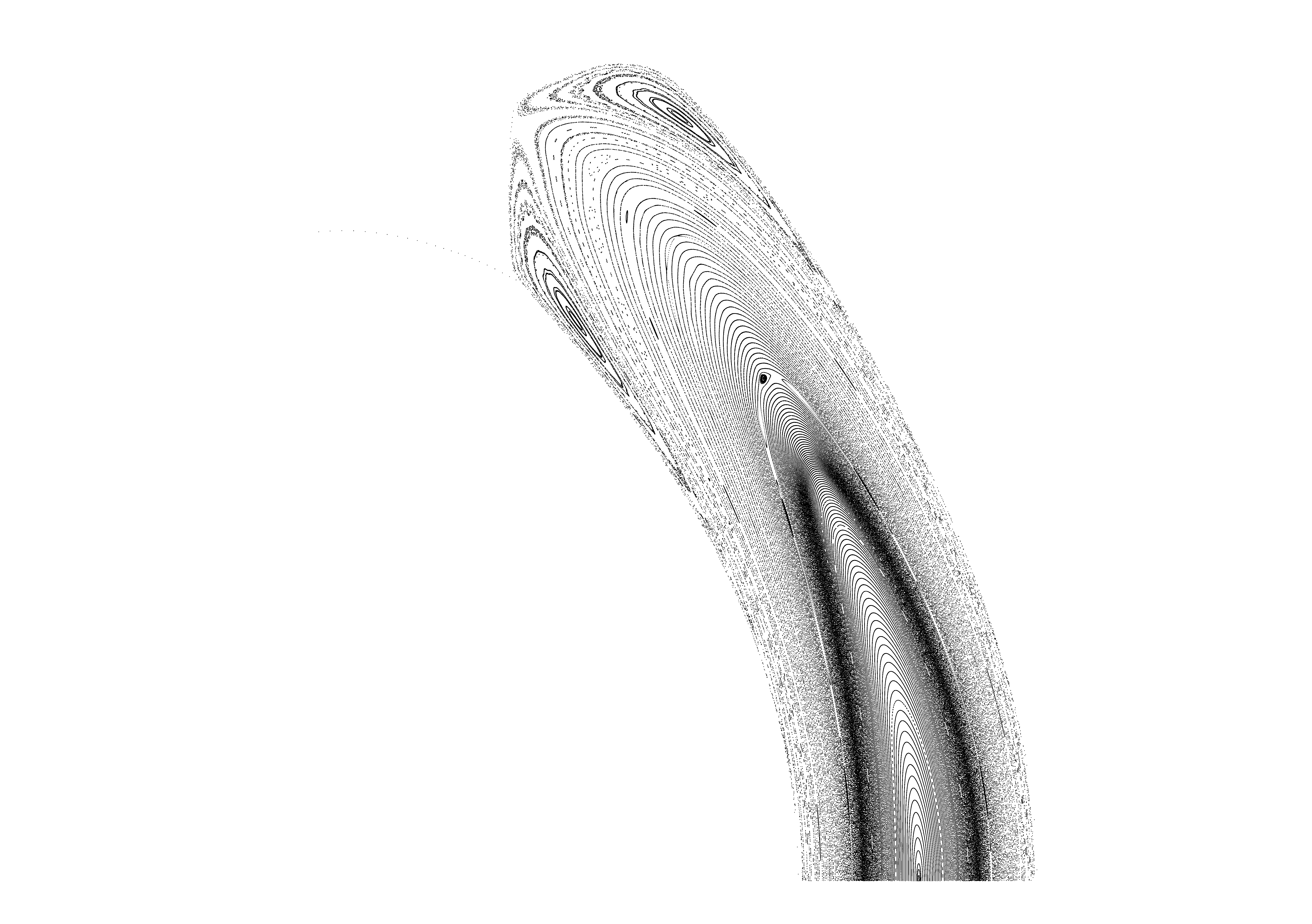}\vspace{0.15cm}
	\includegraphics[draft=false,  trim=315 330 255 365, clip, width=7.5cm]{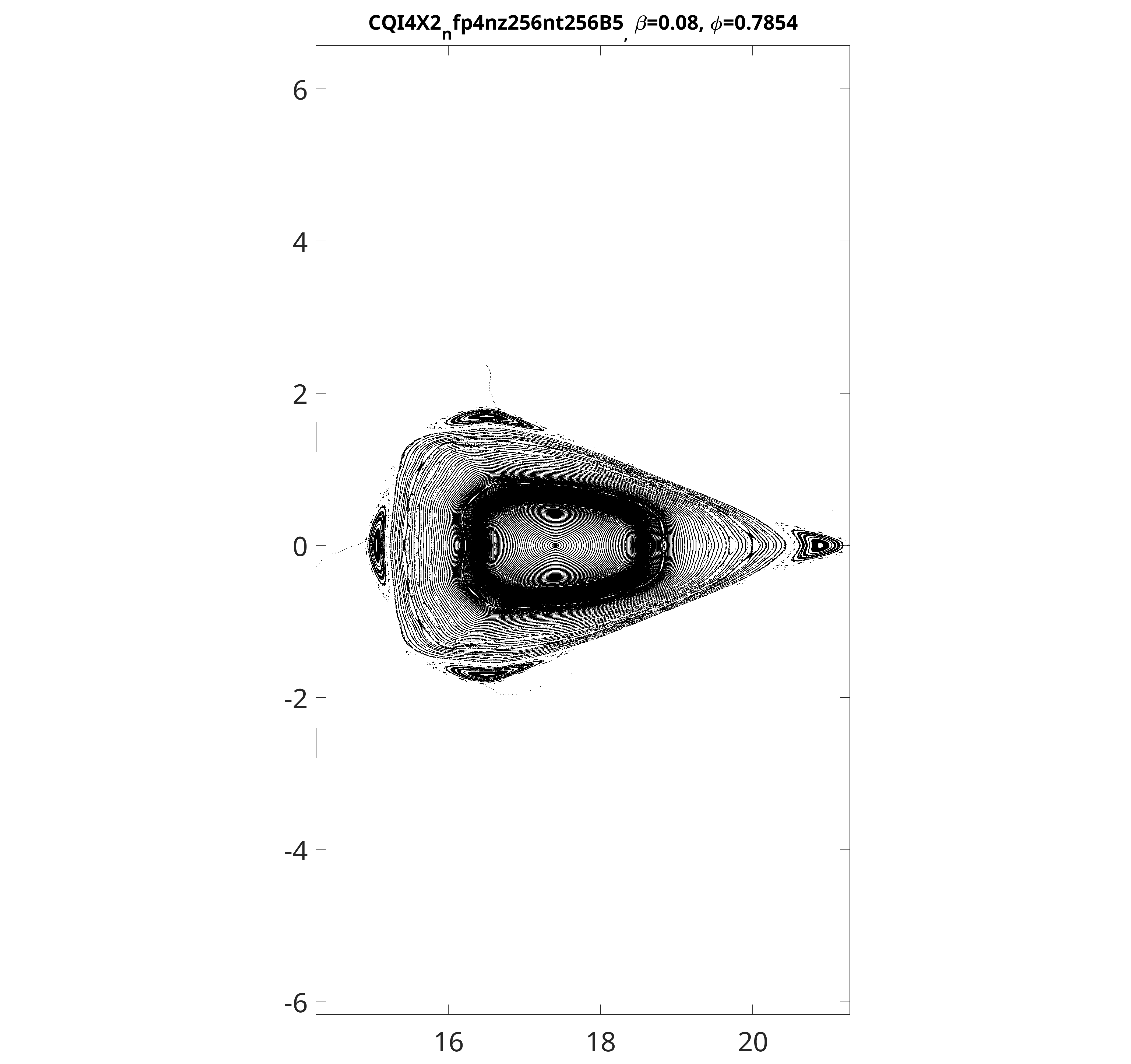}
	\includegraphics[draft=false,  trim=290 100 215 50, clip, width=5.75cm]{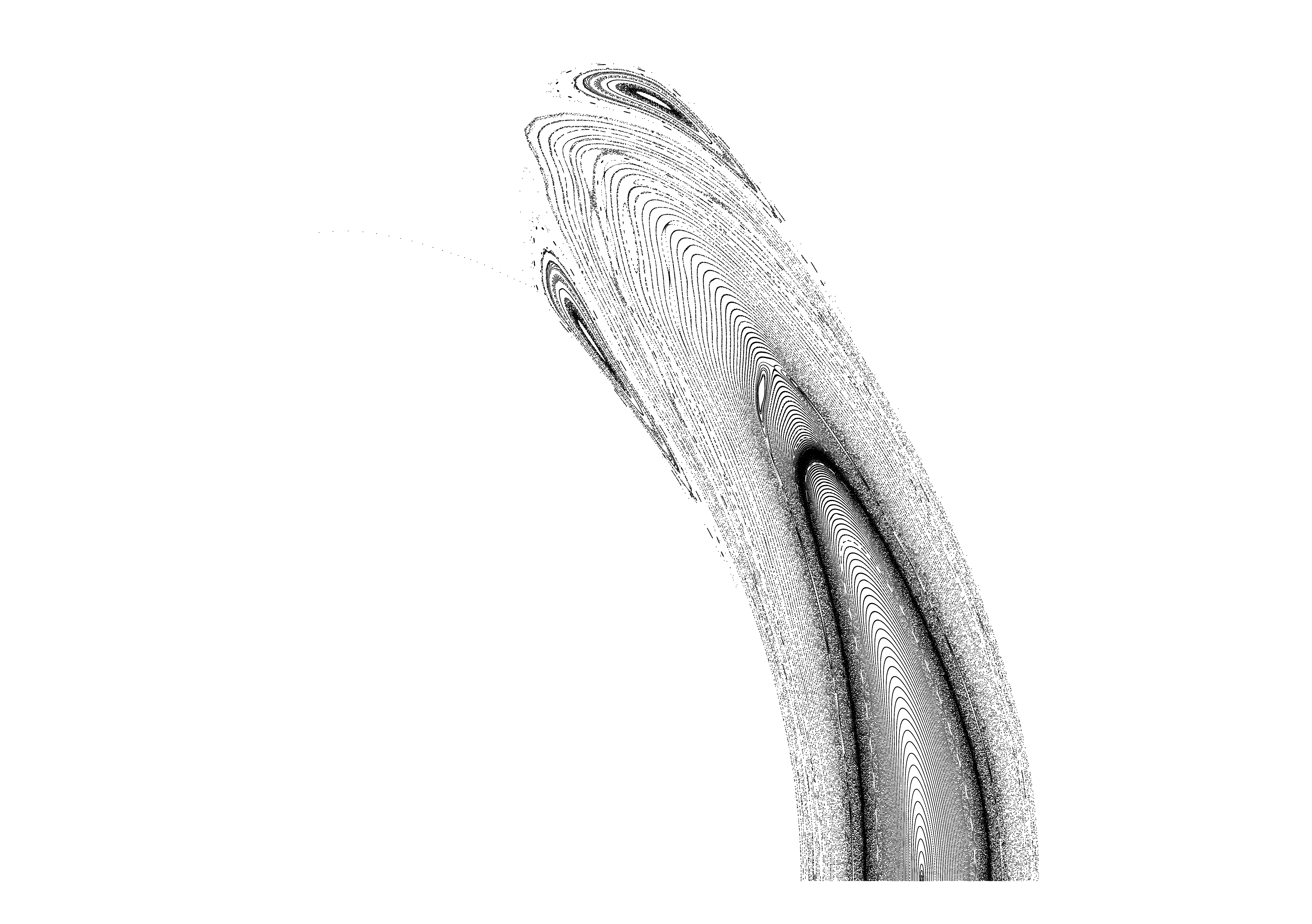}
	\caption{Poincaré plots obtained with HINT-3D for the CIEMAT-QI4X configuration at toroidal angles $\phi=\pi/4$ and $\phi=0$,  and for three values of the plasma pressure, $\beta=0.05\%, 2\%$, and $4\%$.} 
	\label{fig:CQI4XPoincare}
\end{figure*}
		\subsection{Magnetic field structure and flux surfaces}\label{secFSs}

		Figure \ref{fig:CQI4XPoincare} shows Poincaré plots at two toroidal angles, $\phi=\pi/4$ and $\phi=0$, for three plasma pressures, $\beta=0.05\%, ~2\%$, and $4\%$. These equilibria have been computed with the three-dimensional MHD code HINT-3D \cite{suzuki_development_2006}, using the background magnetic field generated by the coils described in Section \ref{secCoils}.  Further details on the HINT-3D calculations can be found in Appendix \ref{apendiceHINT}.  	
		
		In figure \ref{fig:CQI4XPoincare}, the $\tfrac{8}{9}$ island appears at a more inward radial position than in figure \ref{fig:cqi4Poincare}, consistent with the rotational transform profiles in figure \ref{fig:rotationalTransforms}. Its width is also significantly reduced, in agreement with Eq.~(\ref{Eq:islandWidth}), the $\iota$ profiles in figure \ref{fig:rotationalTransforms}, and the discussion in Section \ref{secOptimFSID}. The  quality of the nested flux surfaces between the $\tfrac{8}{9}$ island and the last closed flux surface is substantially improved in figure \ref{fig:CQI4XPoincare}, compared with figure \ref{fig:cqi4Poincare}, and no evidence of overlapping islands or ergodization is observed in this region. The $\tfrac{12}{13}$ island is barely visible in CIEMAT-QI4X (figure \ref{fig:CQI4XPoincare}).

		A well-defined $\tfrac{4}{4}$ island appears at the edge, outside the last closed flux surface, with a substantial radial extent suitable for supporting an island divertor. Its larger size in CIEMAT-QI4X compared with CIEMAT-QI4 is consistent with the lower magnetic shear near the edge (see figure \ref{fig:rotationalTransforms}). As $\beta$ increases, these features are qualitatively preserved. Specifically, the $\tfrac{8}{9}$ island width decreases slightly, the $\tfrac{12}{13}$ island becomes marginally larger, and the edge $\tfrac{4}{4}$ island shrinks somewhat, but its phase remains unchanged—an essential condition for the design of an island divertor with fixed or slightly movable plates.  
		
		\begin{figure}
			\centering
			\includegraphics[draft=false,  trim=50 35 70 25, clip, width=7.75cm]{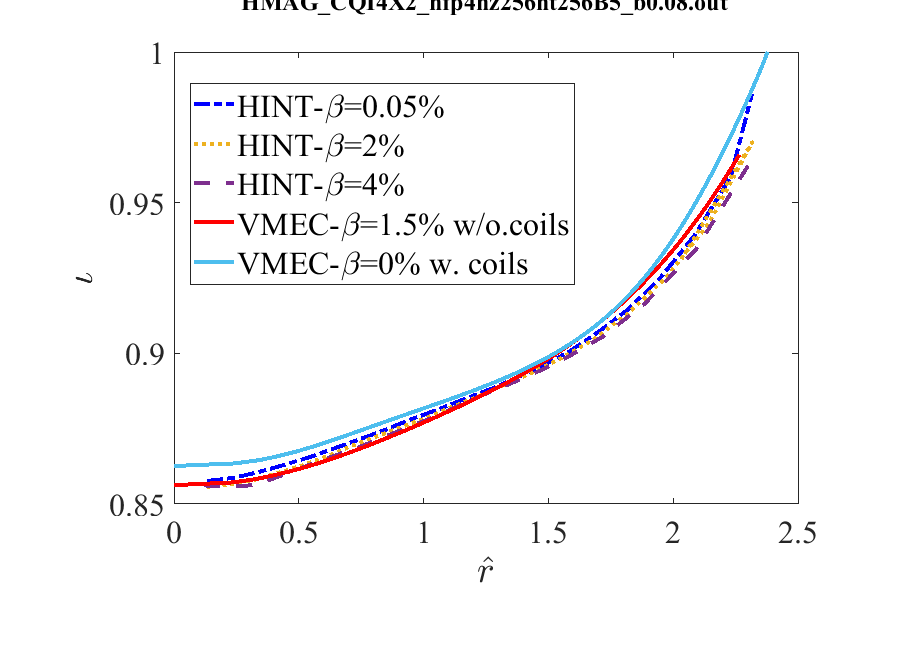}\\
			\caption{Rotational transform profiles for the CIEMAT-QI4X equilibria computed with HINT-3D for different values of $\beta=0.05\%, 2\%, 4\%$, and computed from VMEC fixed-boundary (labeled "w/o. coils") at $\beta=1.5\%$ and vacuum free-boundary calculations (labeled "w. coils"). The coils shown in figure \ref{fig:filamentaryCoils} are used in the VMEC free-boundary and HINT-3D calculations (see the text for the definition of the radial coordinate $\hat{r}$).}
			\label{fig:iotaProfsHINT}
		\end{figure}
	Figure \ref{fig:iotaProfsHINT} compares the rotational transform profiles obtained with HINT-3D for $\beta = 0.05\%$, $~2\%$, and $4\%$ with those from VMEC fixed-boundary calculations at $\beta = 1.5\%$, as well as from vacuum free-boundary VMEC calculations using the filamentary coils described in Section \ref{secCoils}. Since the radial coordinate used in VMEC (normalized toroidal flux) is not well defined in the presence of magnetic islands, we adopt the flux-surface-averaged distance to the magnetic axis, $\hat{r}$, as the radial coordinate for this comparison. In HINT-3D, $\hat{r}$ is computed by integrating along a field line:  		
		\begin{equation}
			\hat{r} = \frac{1}{ \int_L \frac{dl}{B}} \int_L \frac{\sqrt{\left[R(l)-R_{ax}(l)\right]^2 - \left[Z(l)-Z_{ax}(l)\right]^2} dl}{B}, \nonumber
		\end{equation}
		where $(R_{ax}(l),Z_{ax}(l))$ are the $R$ and $Z$ cylindrical coordinates of the magnetic axis. In VMEC, $\hat{r}$ is calculated as the flux-surface average of the distance to the axis,  		
		\begin{equation}
		\hat{r}=\left<\sqrt{\left[R(\theta,\zeta)-R_{ax}(\theta,\zeta)\right ]^2 - \left[Z(\theta,\zeta)-Z_{ax}(\theta,\zeta)\right]^2}\right>.\nonumber
		\end{equation}
		
		The agreement between VMEC fixed-boundary, vacuum free-boundary, and HINT-3D calculations at low $\beta$ is excellent, as expected. Moreover, the HINT-3D rotational transform profile remains essentially unchanged for increasing values of $\beta$ up to $\beta=4\%$.  
		
		The robustness of the inner $\tfrac{8}{9}$ island (see figure \ref{fig:CQI4XPoincare}) follows directly from the near invariance of the $\iota$ profile with $\beta$ shown in figure \ref{fig:iotaProfsHINT}. Similarly, the stability of the outer divertor island with increasing $\beta$ reflects the weak variation of the edge rotational transform. These results confirm the suitability of the CIEMAT-QI4X configuration for operation near the targeted reactor design point at $\beta \sim 3.5\%$ (Section \ref{secConfigProps}).

		\subsection{MHD stability}\label{secNCProps}
		In Ref.~\cite{mercier_necessary_1960}, a necessary condition for stability against ideal interchange instabilities was introduced. The so-called Mercier criterion states that the quantity $D_M$, defined as a sum of several contributions,  

		\begin{equation}
			D_M=D_S+D_I+D_W+D_G,    \\	
			\label{eq:mercierCriterion}		
		\end{equation}
		 must be positive for a configuration to be (locally) stable against ideal interchange modes. Here, $D_S$, $D_I$, $D_W$, and $D_G$ denote the contributions from magnetic shear, plasma current, magnetic well, and geodesic curvature, respectively. The precise definitions of these terms, as implemented in VMEC (used in this work), are given in Appendix~\ref{apendiceMercier}.
		\begin{figure}
			\centering
			\includegraphics[draft=false,  trim=50 10 70 20, clip, width=8.25cm]{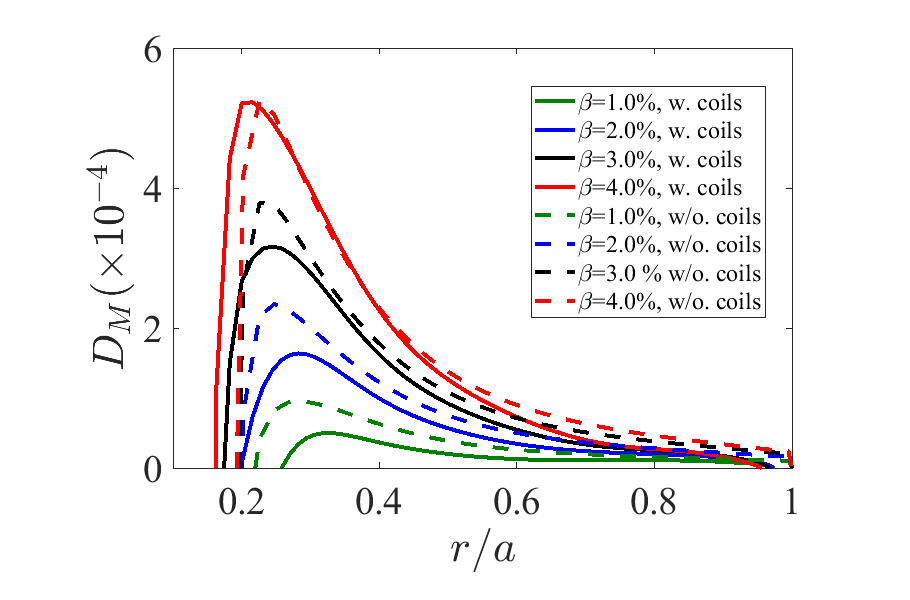}\\
						\caption{$D_M$  from the Mercier criterion  \cite{mercier_necessary_1960} versus radial coordinate, $r/a=\sqrt{s}$, for the CIEMAT-QI4X configuration for several values of $\beta$. $D_M>0$ indicates stability to ideal interchange modes. Values considering the coil ripple  ("w. coils" in the figure) and without coil ripple ("w/o. coils") are shown.}
			\label{fig:mercierDm}
		\end{figure}
		At low $\beta$, the magnetic well contribution $D_W$ usually dominates Eq.~(\ref{eq:mercierCriterion}) \cite{mercier_equilibrium_1964}, which explains why it has often been used as the sole MHD stability criterion in optimization. However, the geodesic curvature term $D_G$ can provide a destabilizing (negative) contribution that increases with $\beta$ \cite{dominguezIdealLownMercier1989a,Varias1990}, potentially driving interchange instabilities, particularly at the plasma edge.  
		
		In some of these recent configurations, magnetohydrodynamic (MHD) stability—although a fundamental requirement for any magnetically confined plasma—was not explicitly considered during optimization \cite{jorge_single-field-period_2022, mata_direct_2022,goodman_constructing_2023}. 
		In other cases, MHD stability was pursued solely through the optimization of the magnetic well as an ideal stability proxy  \cite{anderson_helically_1995,drevlak_optimisation_2019,sanchez_quasi-isodynamic_2023,goodman_constructing_2023}, which is often not sufficient, since the stability of ideal interchange modes depends not only on the magnetic well but also on field line curvature, magnetic shear, and plasma current \cite{mercier_necessary_1960,mercier_equilibrium_1964,landreman_magnetic_2020}.
		
		The CIEMAT-QI4X configuration was optimized at $\beta = 1.5\%$. 
		 During the optimization process, some MHD-unstable configurations were obtained due to the growth of the $D_G$ term. To avoid this, CIEMAT-QI4X optimization has explicitly included $D_M$—in addition to the magnetic well—as a target, ensuring ideal MHD stability. While $D_M$ alone could suffice, magnetic well targets were also applied at many radial positions with wide tolerances to increase the number of constraints in STELLOPT and thereby improve numerical robustness without additional computational cost.
		
		Figure~\ref{fig:mercierDm} shows $D_M$ as a function of normalized radius $r/a$ for CIEMAT-QI4X at several values of $\beta$, both with and without coil ripple. In both cases $D_M$ remains positive across most of the plasma radius
		\footnote{As is well known, $D_M$ computed by VMEC is typically negative at the innermost radial positions.} from $r/a=0.2$ to $r/a=0.95$, and increases with $\beta$ from $\beta=1\%$ to $\beta=4\%$. This indicates that interchange stability is robust and improves with plasma pressure. The magnetic well (not shown) also deepens with increasing $\beta$.  The Mercier criterion, as quantified by $D_M$, is violated at the plasma edge ($r/a>0.95$). 
		These results are very similar to those obtained for recent QI reactor candidate designs \cite{lionStellarisHighfieldQuasiisodynamic2025,hegnaInfinityTwoFusion2025}.

	Although ballooning stability was not explicitly enforced during the optimization, it was assessed a posteriori for high wave-number modes using COBRA \cite{sanchez_cobra_2000}. All ideal ballooning mode growth rates were found to be negative at all radii for $\beta \leq 4\%$ when a small number of helical wells ($k_w \leq 5$) was considered, confirming stability of highly localized (high-$n$) ballooning modes within this pressure range. For the same values of $\beta$, the growth rates increase with the number of helical wells included in the calculation (i.e., approaching the interchange limit), and for $k_w > 10$ they become slightly positive for $r/a > 0.95$, indicating the onset of instability. At higher $\beta$ values, the transition to positive growth rates occurs for smaller $k_w$. This trend is consistent with the negative values of $D_M$ observed in this region. It should be noted, however, that, as suggested in \cite{schmitt_magnetohydrodynamic_2025}, a slight relaxation of profiles at the edge can be sufficient to stabilize ballooning modes.  Furthermore, negative $D_M$ values near the plasma edge—predicting interchange instability—have been experimentally shown not to pose a significant threat in LHD and TJ-II \cite{liang_measurement_2002, ichiguchi_multi-scale_2011, de_aguilera_magnetic_2015}.

		\subsection{Neoclassical transport}\label{secNCProps}
	We now turn to review the neoclassical transport properties of the CIEMAT-QI4X configuration, characterized through the effective ripple, $\epsilon_{\text{eff}}$, which encapsulates the dependence of bulk ion particle and energy transport on the magnetic configuration.  
		
	\begin{figure}
		\centering
		\includegraphics[draft=false,  trim=30 20 60 20, clip, width=8.25cm]{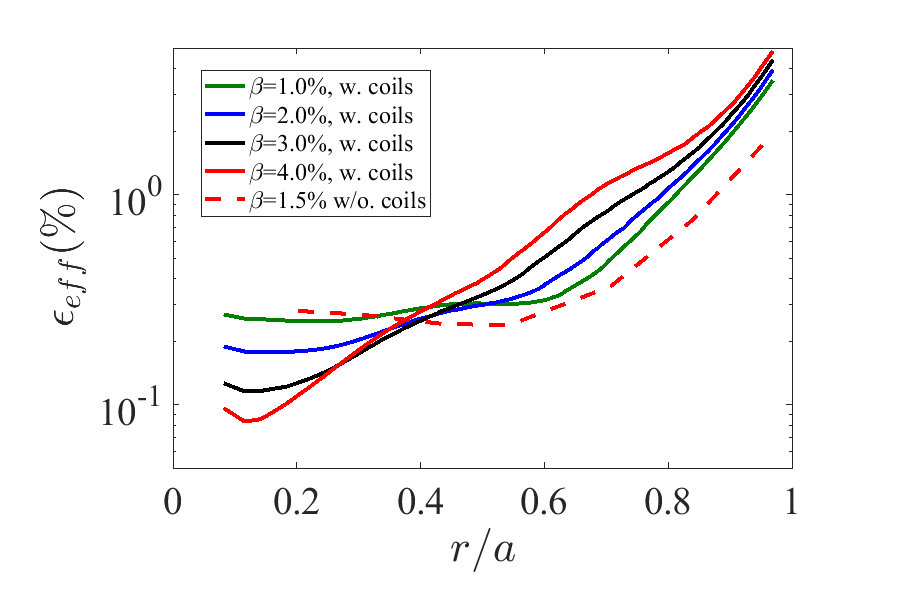}
		\caption{Effective ripple, $\epsilon_{eff}$, versus radial coordinate, $r/a$, for the CIEMAT-QI4X configuration for several values of $\beta =0\%$,  $1.5\%$ and $ 4\%$. Values are shown for the configuration obtained in the optimization (fixed-boundary VMEC calculations without coils) and also for free-boundary VMEC calculations including the coils from Section \ref{secCoils}}
		\label{fig:effRipple}
	\end{figure}
	Figure~\ref{fig:effRipple} shows $\epsilon_{\text{eff}}$ versus normalized radius $r/a$ for  the optimized (fixed-boundary VMEC equilibrium) configuration at $\beta=1.5 \%$ and several free-boundary VMEC equilibria  including the coils described in Section~\ref{secCoils} for different $\beta $ values. 	
	For the ideal configuration, $\epsilon_{\text{eff}}$ is essentially independent of $\beta$. For the free-boundary equilibria including the coil ripple, $\epsilon_{\text{eff}}$ slightly increases with $\beta$ in the plasma edge, but it decreases in the plasma core, up to $0.01\%$ at $r/a<0.2$ for $\beta=4\%$. It is always below $0.5\%$ for $r/a<0.5$. This level is generally considered sufficient for a reactor \cite{beidler_demonstration_2021,Alonso2022}. The main impact of the coils is observed near the plasma edge, where $\epsilon_{\text{eff}}$ increases from about $1.5\%$ up to more than $4\%$ at $\beta=4\%$. It should be noted that the neoclassical transport is usually not the dominant transport mechanism in the plasma edge due to the high collisionality \cite{Dinklage2013}. Furthermore, a large  effective ripple at the edge has been shown to favor the formation of an electron root in the outer half of the plasma, one of the ingredients of the impurity hole \cite{fujitaStudyImpurityHole2021a, velascoModerationNeoclassicalImpurity2016, idaObservationImpurityHole2009}. This is precisely the same mechanisms that has been later proposed for the removal of impurities in newly designed configurations, see e.g. \cite{helander_optimised_2024}.

		\subsection{Bootstrap current}\label{secBootstrap}
\begin{figure}
			\centering
			\includegraphics[draft=false,  trim=40 30 50 30, clip, width=7.75cm]{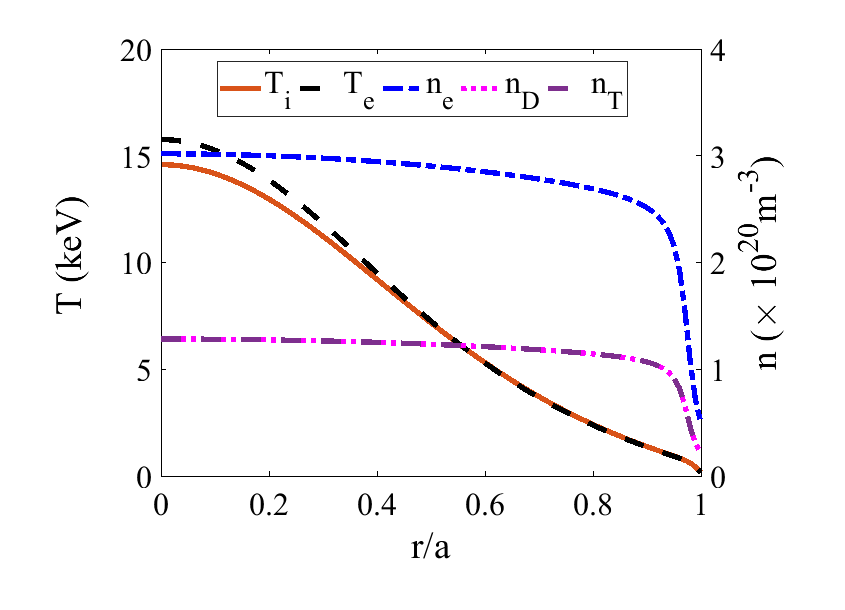}
			\caption{Density and temperature profiles for the reactor design point considered for the bootstrap calculation.}
			\label{fig:reactorProfiles}
		\end{figure}

		In this section we present the results of calculations of the  bootstrap current  for the CIEMAT-QI4X configuration. We consider a reactor with major radius $R=18.5~\rm{m}$, minor radius $a=1.8~\rm{m}$, magnetic field on axis $B_0=5.5~\rm{T}$, and an operating point at $\beta=3.5\%$, corresponding to a fusion power of $P_{\text{fus}}=2.7~\rm{GW}$. The plasma consists of a 50/50 D–T mixture, electrons, and fusion-born alpha particles, with the density and temperature profiles shown in Fig.~\ref{fig:reactorProfiles} \cite{Alonso2022}.  
			
\begin{figure}
			\centering
			\includegraphics[draft=false,  trim=70 50 80 30, clip, width=7.75cm]{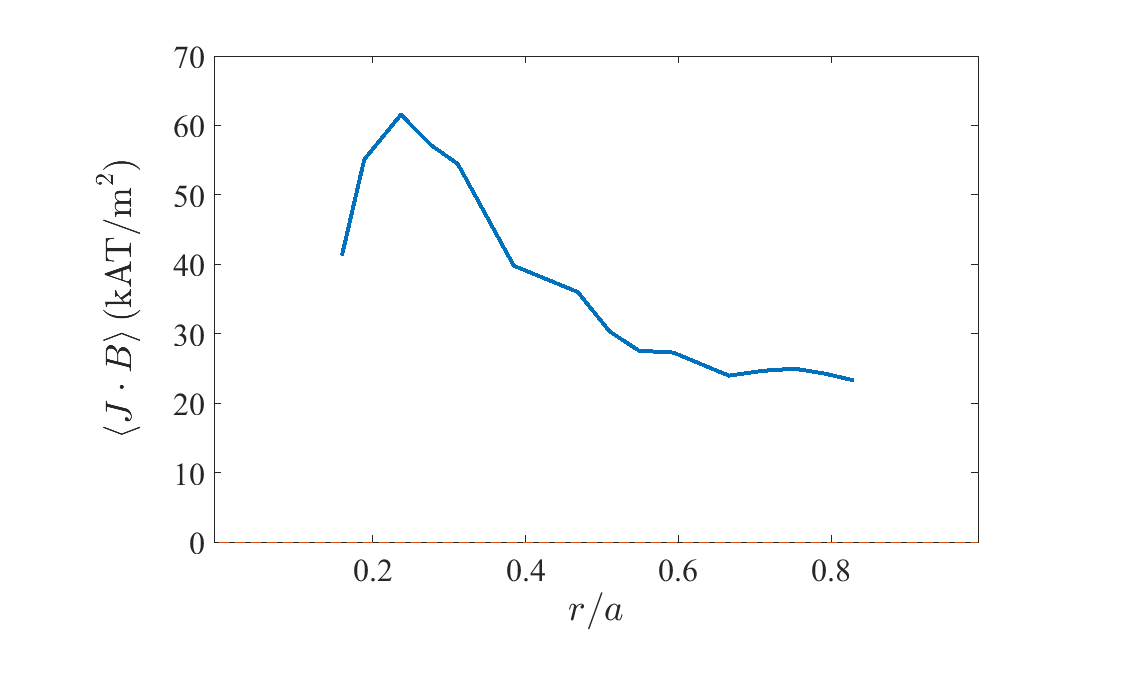}
			\caption{Flux surface average of the parallel boostrap current computed with SFINCS \cite{landremanOmnigenityGeneralizedQuasisymmetrya2012} for  CIEMAT-QI4X, using the density and temperature profile from figure \ref{fig:reactorProfiles}}
			\label{fig:BSJdotrBFSAve}
		\end{figure}

\begin{figure}
			\centering
			\includegraphics[draft=false,  trim=615 270 555 270, clip, width=8.25cm]{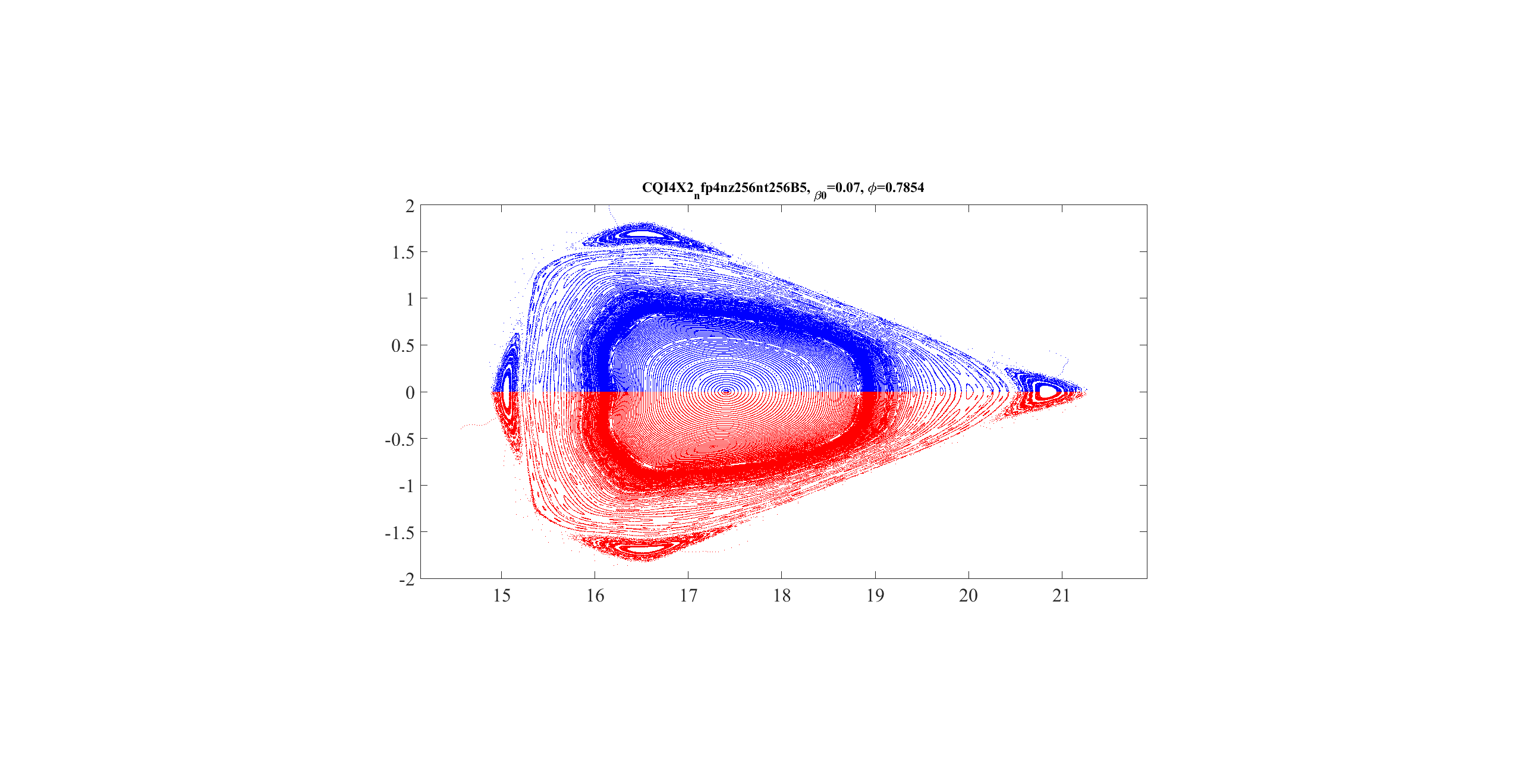}
			\caption{Poincaré plots for two toroidal angle $\phi=0$ obtained from HINT-3D calculations for the CIEMAT-QI4X configuration at $\beta=3.5\%$ with (blue) and without (red) accounting for the neoclassical bootstrap current shown in figure \ref{fig:BSJdotrBFSAve}.}
			\label{fig:HINTwBS}
		\end{figure}

		The bootstrap current was computed with SFINCS \cite{landremanOmnigenityGeneralizedQuasisymmetrya2012} using these profiles. Figure~\ref{fig:BSJdotrBFSAve} shows the flux-surface-averaged parallel bootstrap current density obtained from these calculations versus normalized radius. Integrating over the plasma volume yields a total toroidal current of $54~\rm{kA}$. When this current is included in the VMEC input and free boundary calculations are performed, the resulting change in the rotational transform is smaller than $2.2\%$ throughout the plasma and about $1\%$ at the edge, relative to the current-free case (not shown).
		The impact of the bootstrap current was also tested with HINT-3D free-boundary calculations using the coil set of Fig.~\ref{fig:filamentaryCoils}. Including the bootstrap current does not significantly modify the rotational transform profile or the flux surfaces. Figure~\ref{fig:HINTwBS} compares Poincaré plots at $\phi=\pi/4$ for calculations with and without the bootstrap current, showing negligible differences.  
		
		The small bootstrap current obtained under reactor conditions ensures the stability of the rotational transform profile, the nested flux surfaces, and, crucially, the divertor island structure, thus supporting the viability of an island divertor in this configuration.
		
		\subsection{{Fast ion confinement}}\label{FIMetrics}
		In this section we study the confinement of fusion $\alpha$-particles using collisional calculations with ASCOT \cite{akaslompolo_validating_2019}, for a configuration scaled to reactor size and magnetic field. We adopt the same density and temperature profiles as in \cite{landreman_optimization_2022}, with minor radius $a=1.7~\rm{m}$ and magnetic field on axis $B_0=5.7~\rm{T}$. Fusion-born $\alpha$-particles are followed for times longer than the slowing-down time, and the fraction of particles and energy that are lost is computed.  
		
\begin{figure}
		 	\centering
		 	\includegraphics[draft=false,  trim=30 30 10 10, clip, width=8.5cm]{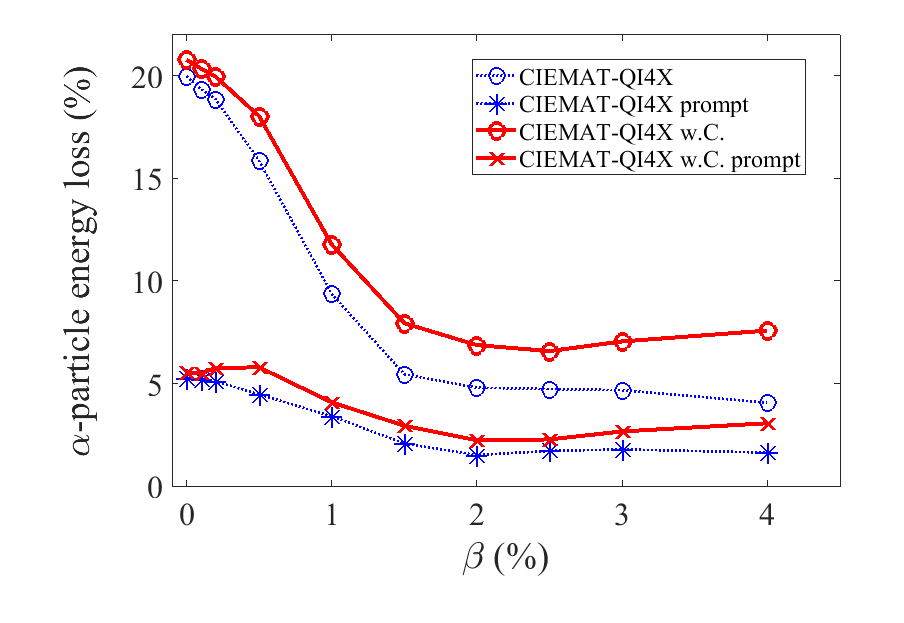}
		 	\caption{Fraction of energy from $\alpha$ particles e`scaping from the plasma versus plasma $\beta$. The calculations are done with ASCOT using the profiles from figure \ref{fig:reactorProfiles}. The fraction of energy lost after a time longer than slowing-down time are shown for CIEMAT-QI4X including coils (labeled 'w.C.'), and without coils. The fraction of energy lost by particles before thermalization is shown separately ('x' and '*' symbols, labeled as 'prompt').}
		 	\label{fig:fastIonConfinement}
		 \end{figure}

		Calculations are performed for a set of  fixed-boundary equilibria obtained with VMEC  at different plasma pressures up to $\beta=4\%$, and also for a set of free-boundary VMEC equilibria including the coils of Sec.~\ref{secCoils}.  
		Figure~\ref{fig:fastIonConfinement} shows the fraction of $\alpha$-particle energy lost after thermalization time for CIEMAT-QI4X (with and without coil ripple). Prompt losses are also shown separately.

		 The energy losses decrease significantly with $\beta$ increasing from vacuum up to $\beta=1.5\%$ (the optimization point for CIEMAT-QI4X), and remain nearly constant from $\beta=1.5~\%$ up to $\beta=4\%$. The confinement of energetic particles is very good in both sets (with and without coil ripple) of configurations, but coil ripple has a noticeable effect, increasing losses by up to $40\%$ for some values of $\beta$. Nevertheless, the total fraction of energy lost, including coil ripple, remains below $8\%$ for $2\%<\beta<4\%$, which is considered acceptable from the point of view of the $\alpha$-heating efficiency (typically assumed $\sim 0.9$ \cite{sagaraReviewStellaratorHeliotron2010a}) and consistent with reactor scenarios in \cite{Alonso2022}.  
		 
		 The prompt losses, which are the most critical from the perspective of potential first-wall damage, remain below $3\%$ for $2\%<\beta<4\%$. 
		  A detailed assessment of the implications of these losses requires dedicated studies that include not only the magnetic configuration and angular distribution of escaping particles, but also engineering aspects such as coil design, shielding, divertor, and blanket — tasks beyond the scope of this work.  { 
		  	At high $\beta$, the free-boundary VMEC equilibria differ substantially from HINT results, especially in the position of the magnetic axis and the shape of the plasma boundary, which expands unrealistically outward relative to HINT-3D. Consequently, calculations that include the coils at high $\beta$ likely overestimate the fast-ion energy losses.}

		\subsection{Turbulent transport}\label{secTurbTransp}
		\begin{figure}
			\centering
			\includegraphics[draft=false,  trim=0 50 50 105, clip, width=7.25cm]{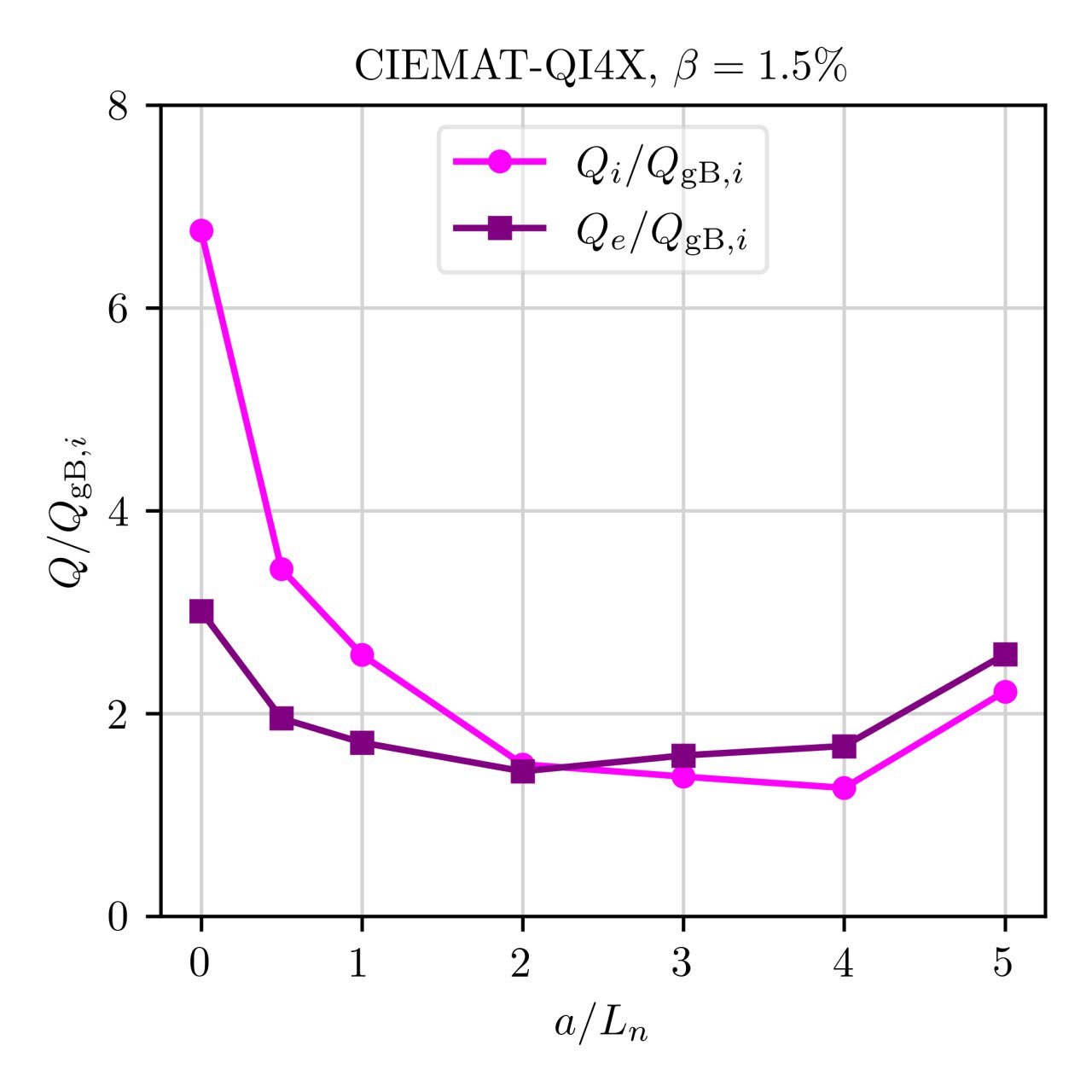}\\
			\includegraphics[draft=false,  trim=50 50 20 105, clip, width=7.75cm]{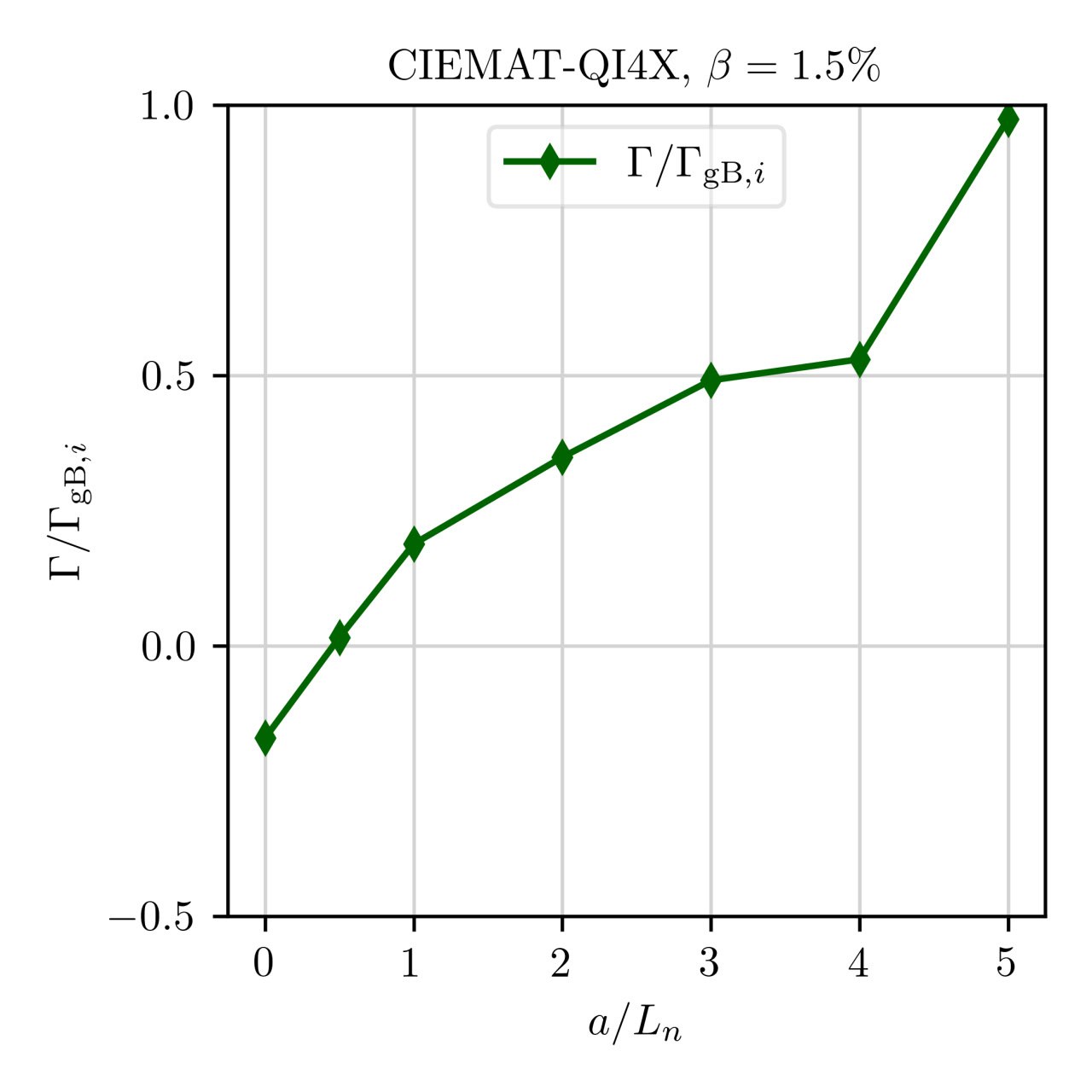}\\
			\caption{Ion and electron heat fluxes, $Q_i$ and $Q_e$, and particle flux, $\Gamma$, (normalized to gyro-Bohm units) versus normalized density gradient, $a/L_n$  for the CIEMAT-QI4X configuration at $\beta=1.5\%$. In all cases, $a/L_{T_{e}}=a/L_{T_{i}}=3$}
			\label{fig:qiGB}
		\end{figure}	
		Regarding turbulent transport, as for CIEMAT-QI4 \cite{garcia-regana_reduced_2024}, we have evaluated the electrostatic turbulent transport in CIEMAT-QI4X using radially local flux-tube simulations with the code \textsc{stella}. We consider an MHD equilibrium with $\beta=1.5\%$ (the optimization value for this configuration) and include both kinetic ions and electrons. Simulations are performed at $r/a=0.7$ in a flux tube defined by $\alpha=0$, with $\alpha=\theta^*-\iota\zeta$, extending approximately two poloidal turns. Here, $\theta^*$ and $\zeta$ are the poloidal and toroidal PEST coordinates.
	
		Wavenumbers in the range $0.063 \rho_i ^{-1}< k_{x} < 3 \rho_i ^{-1}$ and $0.067 \rho_i ^{-1}< k_y < 2\rho_i ^{-1}$ are resolved with grid resolutions $n_x \times n_y \times n_z = 135 \times 92 \times 256$. The thermal ion Larmor radius is $\rho_i = v_{th,i}/\Omega_i$, where $v_{th,i}=\sqrt{2T_i/m_i}$ and $\Omega_i=eB_r/m_i$ is the ion gyrofrequency, with $e$ the electron charge, $m_i$ the ion mass, $T_i$ the ion temperature at $r/a=0.7$, and $B_r$ a reference magnetic field strength \cite{Barnes19,Gonzalez-Jerez}. The velocity-space resolution is $N_{\mu}=12$ for the magnetic moment and $N_{v_{\|}}=64$ for the parallel velocity coordinate $v_{\|}$.  
		
		A scan in the density gradient is carried out, from $a/L_n=0$ to $a/L_n=5$, while keeping the ion and electron temperature gradients fixed at $a/L_{T_i}=a/L_{T_e}=3$, which are typical of experimental plasmas in W7-X. Here, $a/L_X=-(a/X)(dX/dr)$ denotes the normalized gradient scale length of quantity $X$.  
		
		The results of this density-gradient scan are shown in Fig.~\ref{fig:qiGB}, which presents the gyro-Bohm-normalized ion and electron heat fluxes and the particle flux versus $a/L_n$. At $a/L_n=0$, the ion heat flux is approximately twice the electron heat flux. Increasing the density gradient from 0 to 2 leads to a strong reduction of both fluxes: $Q_i$ decreases by about a factor 4.5, and $Q_e$ by a factor 2. For $2<a/L_n<4$, a regime of very low heat transport appears, with $1.3 \lesssim Q_i/Q_{i,GB} \approx Q_e/Q_{i,GB} \lesssim 2$.  This reduction of heat fluxes with density gradient has already been observed in W7-X and CIEMAT-QI4 \cite{garcia-regana_reduced_2024,hanne23} and has been experimentally shown to allow the access to an enhanced performance regime in W7-X \cite{bozhenkov_high-performance_2020}. This turbulence stabilization by density gradient is larger in CIEMAT-QI4X (and CIEMAT-QI4) than in W7-X (see \cite{garcia-regana_reduced_2024} for comparison). As for CIEMAT-QI4, this reduction is attributed to the approach to the maximum-$J$ property, which is an optimization criterion, in this configuration.  
		
		For larger density gradients ($a/L_n>4$), both ion and electron heat fluxes increase only slightly, supporting the robustness of the configuration to increasing density gradients under experimentally relevant conditions.  
		
		The particle flux $\Gamma$ is slightly negative (turbulent pinch) at $a/L_n=0$, which has been shown to counteract the neoclassical thermodiffusion responsible for density-core depletion in stellarators \cite{hanne23}. For $1<a/L_n<4$, $\Gamma$ increases up to $\Gamma/\Gamma_{i,GB}\approx0.5$, and reaches $\Gamma/\Gamma_{i,GB}\approx1$ at $a/L_n=5$. This moderate growth supports the formation of a density pedestal, which is considered instrumental for access to high-confinement modes in W7-X \cite{thienpondt_influence_2024, bozhenkov_high-performance_2020}.  
		
		Overall, these results are qualitatively similar to those of CIEMAT-QI4, and indicate improved turbulent-transport properties compared to W7-X \cite{garcia-regana_reduced_2024}. This improvement is interpreted as a consequence of the approach to the maximum-$J$ property already at moderate $\beta$ in the CIEMAT-QI4 and CIEMAT-QI4X configurations.  
		
\section{Summary and conclusions}\label{secSumandConc}
In this work, we present a new four-period quasi-isodynamic stellarator configuration, named CIEMAT-QI4X, which fulfills a number of physics requirements for a stellarator reactor design. 
Its salient feature, compared with the previously reported CIEMAT-QI4 configuration \cite{sanchez_quasi-isodynamic_2023}, is the improvement of flux-surface quality and divertor island structure, while retaining essentially all the favorable physics properties of the CIEMAT-QI4  design. This improvement has been achieved through a strict control of the rotational transform and its magnetic shear during the optimization process. 

A set of optimized filamentary coils has been designed to reproduce this configuration with sufficient fidelity to preserve its physics properties, while at the same time meeting general engineering requirements. According to preliminary assessments of breeding blanket integration for the CIEMAT-QI family of configurations, the coil–plasma separation in this coil set would be sufficient to accommodate a blanket with adequate tritium breeding ratio. The coil–coil separation has been improved with respect to previous designs \cite{sanchez_quasi-isodynamic_2023}, reaching values comparable to other recently reported stellarator reactor studies \cite{lionStellarisHighfieldQuasiisodynamic2025,hegnaInfinityTwoFusion2025}. The coils are relatively simple, with curvature and torsion values considered acceptable in light of earlier design studies \cite{lionStellarisHighfieldQuasiisodynamic2025,hegnaInfinityTwoFusion2025}. Detailed feasibility studies with different conductor technologies, including low-temperature (LTS) and high-temperature superconductors (HTS), are ongoing and will be reported elsewhere. 

The quality of flux surfaces and the divertor island structure have been demonstrated with HINT-3D calculations at several values of the plasma pressure in the range of realistic reactor scenarios, ($\beta<4\%$), 
using the filamentary coil set. Good flux surfaces are obtained in the plasma column, with limited width of the lowest-order islands (e.g. $\frac{8}{9}$) for $0<\beta<4\%$. A well-defined $\frac{4}{4}$ island appears at the edge, whose size and phase are preserved as $\beta$ increases, supporting the feasibility of an island-divertor concept. 

The configuration shows favorable MHD stability properties. Mercier stability improves with plasma pressure up to reactor-relevant values $\beta \sim 4\%$. Ballooning stability, evaluated with the COBRA code, indicates stability against short-wavelength ballooning modes up to $\beta=4\%$. Neoclassical transport is low, improved with respect to W7-X in the plasma core, and comparable to CIEMAT-QI4 \cite{sanchez_quasi-isodynamic_2023}. The effective ripple $\epsilon_{eff}$ remains below $0.5\%$ in the core ($r/a<0.5$), even accounting for coil ripple. The bootstrap current, evaluated with SFINCS for reactor-relevant profiles at $\beta \sim 3.5\%$, gives a total toroidal current of $\sim 54$~kA, sufficiently small to ensure stability of the rotational-transform profile and flux-surface quality. Free-boundary VMEC equilibria including this current show only small changes in the rotational  transform ($<2.25\%$ across the radius), and HINT-3D calculations confirm negligible effects on the rotational transform profile, the flux surfaces and the divertor island.  

The confinement of fast ions improves with $\beta$ and is already very good at $\beta=1.5\%$. The fraction of $\alpha$-particle energy lost after thermalization is below $8\%$ for $2\%\leq \beta \leq 4\%$, consistent with reactor requirements for plasma heating efficiency (commonly assumed $\gtrsim 0.9$). The energy losses from promptly lost $\alpha$-particles remain below $3\%$ in this range of $\beta$.  Studies of energy loads to the wall are underway.

The turbulent transport in this configuration has been analyzed using electrostatic flux-tube simulations with \textsc{stella}, including kinetic ions and electrons. The calculations consider experimentally relevant values of density and temperature gradients. 
The results are comparable to those of CIEMAT-QI4 \cite{garcia-regana_reduced_2024}, showing reduced turbulent transport as compared to W7-X in the parameter range studied. A scan in the density gradient indicates that both ion and electron heat fluxes are low at small density gradient 
and are strongly reduced as it 
 increases to moderate values. 
 For large density gradients, the heat fluxes rise only slightly. 
The reduction of heat fluxes with density gradient, already observed in W7-X and CIEMAT-QI4 \cite{garcia-regana_reduced_2024,hanne23}, supports the access to enhanced performance regimes \cite{bozhenkov_high-performance_2020}.  
The particle flux is slightly negative (turbulent pinch) for a flat density profile, 
and increases moderately with increasing density gradients, 
 supporting the formation of a density pedestal.  

Overall, CIEMAT-QI4X fulfills the main requirements to serve as a physics basis for stellarator reactor design, and further work is underway in this direction. Detailed electro-mechanical studies of the coils using different conductor technologies are in progress, as well as assessments of breeding-blanket integration. Building on the CIEMAT-QI4X optimization process, in which improvements in flux surfaces and divertor structure were achieved by tailoring the rotational transform profile, new metrics have been implemented in the STELLOPT optimization suite. Configurations obtained using these new metrics will be reported in future work.

\section*{Acknowledgments}

E.S is grateful to C. Zhu for providing and supporting the code FOCUS, to C. Zhu and T. Krugger for providing special versions of FOCUS for island correction, to Y. Suzuki for providing and supporting the code HINT-3D and to R. Sánchez for the support with the code COBRA. 
This work has been carried out within the framework of the EUROfusion Consortium,
funded by the European Union via the Euratom Research and Training Programme (Grant Agreement No 101052200-EUROfusion). Views and opinions expressed are however those of the authors only and do not necessarily reflect those of the European Union or the European Commission. Neither the European Union nor the European Commission can be held responsible for them. 
This work has been partially funded by the Ministerio de Ciencia, Innovaci\'on y Universidades of Spain  under projects PGC2018-095307-B-I00 and PID2021-123175NB-I00. Calculations for this work made use of computational resources at XULA (CIEMAT), Mare Nostrum-V and the EUROfusion supercomputers Marconi and Pitagora. We akcnowledge the resources provided by CIEMAT, EUROfusion and the Spanish Supercomputing Network.

\section*{Apendix A. Mercier criterion}\label{apendiceMercier}
In Ref.~\cite{mercier_necessary_1960}, a necessary condition for the stability of ideal interchange instabilities was derived. The so-called Mercier criterion states that the quantity $D_M$, defined as
\begin{equation}
	D_M = D_S + D_I + D_W + D_G,  	
	\label{eq:mercierCriterion}		
\end{equation}
must be positive for the configuration to be (locally) stable against ideal interchange modes. The contributions $D_S$, $D_I$, $D_W$, and $D_G$ are associated with the magnetic shear, toroidal current, magnetic well, and geodesic curvature, respectively. In VMEC, these terms are computed as
\begin{eqnarray}			 	
	D_S &=& \left(\frac{d \iota}{d \psi}\right)^2, \nonumber\\
	D_I &=& \frac{d\iota}{d\psi}\left(\left<\frac{B^2 }{|\nabla \psi|^3}\right> I' - \left<\frac{\mathbf{J}\cdot \mathbf{B}}{|\nabla \psi|^3}\right>\right), \nonumber\\
	D_W &=& \frac{dp}{d\psi}\left(\frac{d^2V}{d\psi^2}- \frac{ dp}{d\psi}\left<\frac{1}{B^2}\right>\right)\left< \frac{B^2}{|\nabla \psi|^3}\right>, \nonumber\\
	D_G &=& \left(\frac{\left< \mathbf{J}\cdot \mathbf{B}\right> }{|\nabla \psi|^3}\right)^2  
	- \left< \frac{B^2}{|\nabla \psi|^3} \right>  
	\left< \frac{(\mathbf{J}\cdot \mathbf{B})^2 }{B^2|\nabla \psi|^3} \right>, \nonumber
	\label{eq:DmTerms}		
\end{eqnarray}
where $\psi$ is the toroidal magnetic flux (used as radial coordinate), $p$ is the plasma pressure, $\mathbf{J}$ the plasma current, $\mathbf{B}$ the magnetic field, $I$ the total toroidal current, and $V$ the plasma volume.  

The term $D_S$, associated with magnetic shear, is always positive and therefore always stabilizing. It is exploited as the main stabilizing mechanism in devices such as heliotrons \cite{Uro1982,wakatani_mhd_1991}, and in combination with $D_W$ in torsatrons \cite{okamura_mhd_1995,harris_second_1989,pavlichenko_first_1993}. The contribution $D_I$, linked to toroidal plasma current, is typically small in stellarators due to their nearly current-free operation. The magnetic-well term $D_W$ can be stabilizing or destabilizing; it is the main stabilizing effect in low-shear devices \cite{hirsch_major_2008,Varias1990,grieger_modular_1992}, or acts in combination with $D_S$ in torsatron/heliotron designs. Finally, the geodesic-curvature term $D_G$ can take either sign; in some configurations it becomes increasingly negative near the plasma edge as pressure rises \cite{Varias1990}.  

All terms in $D_M$ are evaluated by VMEC within the STELLOPT optimization suite. 

\begin{figure*}[!ht]
	\centering
	\includegraphics[draft=false,  trim=10 75 20 25, clip, width=7.75cm]{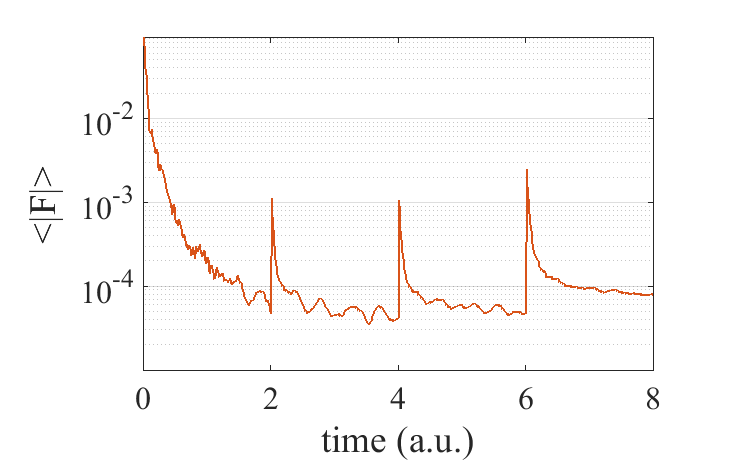}
	\includegraphics[draft=false,  trim=10 75 20 25, clip, width=7.75cm]{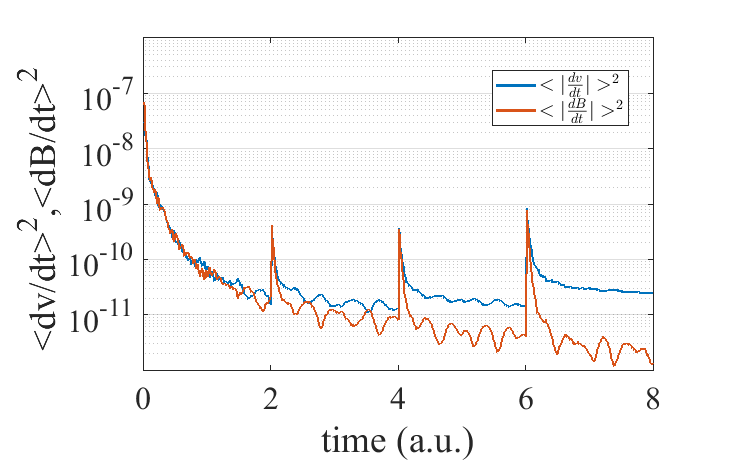}\\
	\includegraphics[draft=false,  trim=10 75 20 25, clip, width=7.75cm]{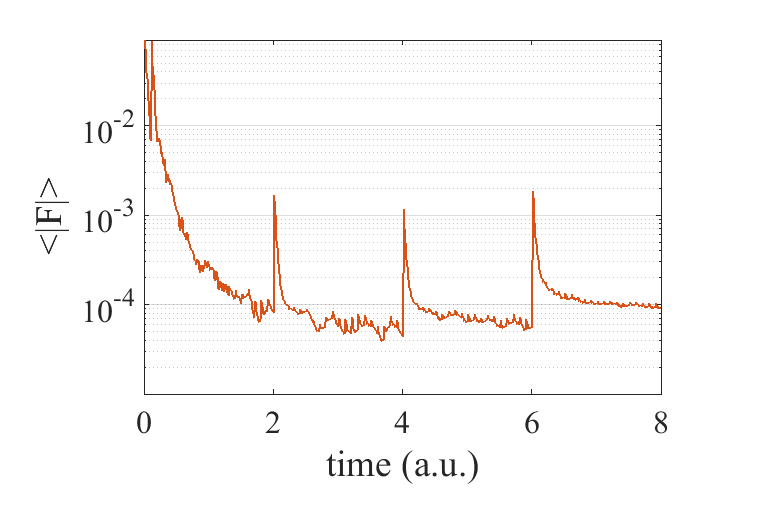}
	\includegraphics[draft=false,  trim=10 75 20 25, clip, width=7.75cm]{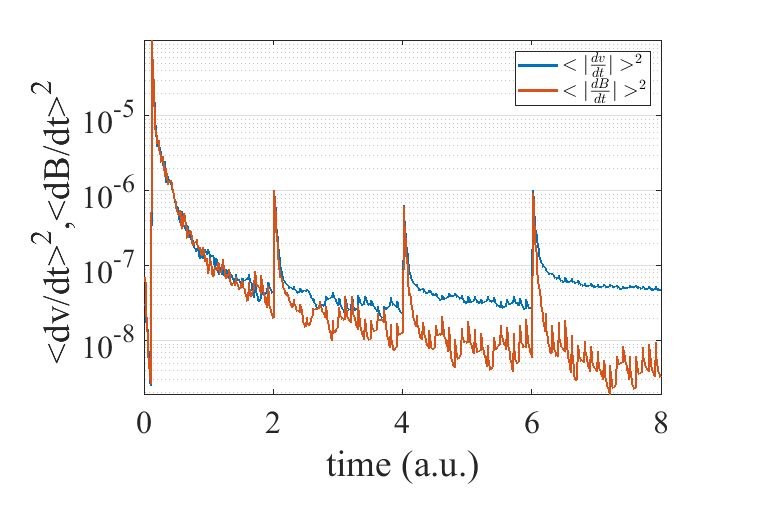}\\
	\includegraphics[draft=false,  trim=10 5 20 25, clip, width=7.75cm]{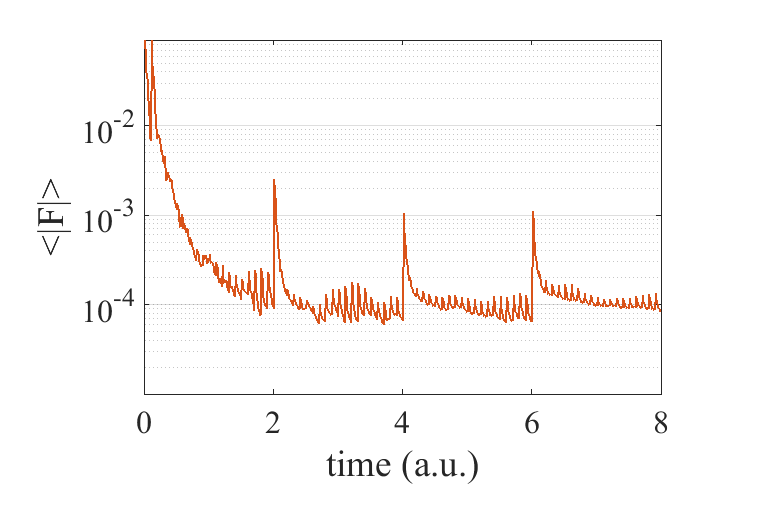}
	\includegraphics[draft=false,  trim=10 5 20 25, clip, width=7.75cm]{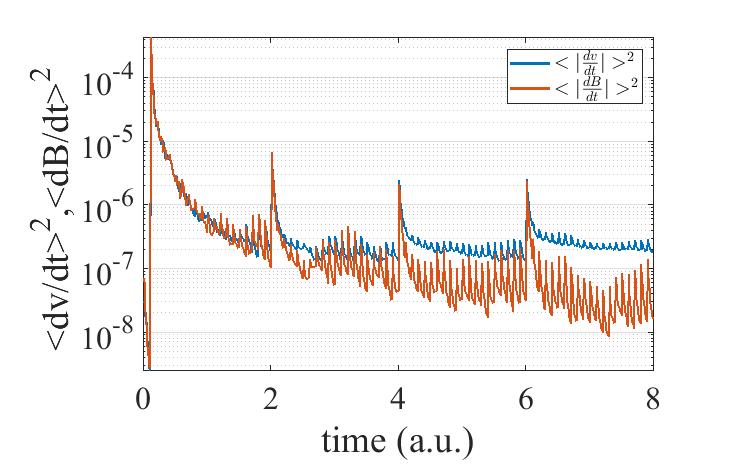}\\
	\caption{Convergence of the HINT-3D calculations for the equilibria at $\beta=0.05\%$ (top row), $2\%$ (middle row) and $4\%$ (bottom row) shown in section \ref{secConfigProps}. }
	\label{fig:HINTconvergence}
\end{figure*}
\section*{Apendix B. HINT-3D calculations details}\label{apendiceHINT}
The HINT-3D calculations for both CIEMAT-QI4 and CIEMAT-QI4X were carried out with high resolution in cylindrical coordinates $R,Z,\phi$, using $n_R\times n_Z\times n_{\phi}=256\times256\times256$. A small resistivity, $\eta_0=10^{-6}$, was employed to ensure convergence to a small residual force. 
Convergence was monitored through the volume-averaged residual force $\left<|F|\right>$, as well as $\left<|\frac{d\mathbf{v}}{dt}|^2\right>$ and $\left<|\frac{d\mathbf{B}}{dt}|^2\right>$ \cite{suzuki_development_2006}. Their evolution during the calculation is shown in Fig.~\ref{fig:HINTconvergence} for the CIEMAT-QI4X equilibria at different values of $\beta$ from Sec.~\ref{secConfigProps}. 

The surface-averaged residual force converges to $\left<|F|\right>\approx 10^{-4}$ for all $\beta$ values considered. For $\beta=4\%$, $\left<|\frac{d\mathbf{v}}{dt}|^2\right>$ and $\left<|\frac{d \mathbf{B}}{dt}|^2\right>$ are in the ranges $10^{-7}$–$10^{-8}$, while for smaller $\beta$ the convergence in this metrics is even better; at $\beta=1\%$, both quantities are several orders of magnitude smaller. Overall, values of $\left<|F|\right>\approx 10^{-4}$, $\left<|\frac{d\mathbf{v}}{dt}|^2\right>\approx 10^{-11}$–$10^{-7}$, and $\left<|\frac{d\mathbf{B}}{dt}|^2\right>\approx 10^{-12}$–$10^{-8}$ are considered sufficient for convergence \cite{suzuki_development_2006}. 

A net plasma current in the range $1.7\,\mathrm{kA}$ (for $\beta=0.05\%$) to $120\,\mathrm{kA}$ (for $\beta=4\%$) appears, which is two orders of magnitude smaller than the coil current ($\sim 12\,\mathrm{MA}$) for $\beta=4\%$ and negligible at low $\beta$. This plasma current has very little impact on the rotational-transform profile, which is only slightly modified, as shown in Fig.~\ref{fig:iotaProfsHINT}. 
%
\section*{}
\bibliography{Sanchez_CQI4X}

\end{document}